\documentclass{aa}

\usepackage{graphicx}
\usepackage{txfonts}
\usepackage{siunitx}
\usepackage{threeparttable}  
\usepackage{color}

\begin{document}

   \title{Connecting solar flare hard X-ray spectra to in situ electron spectra}

   \subtitle{A comparison of RHESSI and STEREO/SEPT observations}

   \author{N. Dresing
          \inst{1,7}
          \and
          A. Warmuth \inst{2}\and
          F. Effenberger \inst{3,4}\and
          K.-L. Klein \inst{5}\and
           S. Musset \inst{6,7}\and
          L. Glesener \inst{6}\and
          M. Brüdern \inst{8}
          }

   \institute{Department of Physics and Astronomy, University of Turku, Finland\\
    \email{dresing@physik.uni-kiel.de}
          \and
         Leibniz-Institut f\"ur Astrophysik Potsdam (AIP), An der Sternwarte 16, 14482 Potsdam, Germany \\
             \email{awarmuth@aip.de}
            \and Institut f\"ur Theoretische Physik, IV, Ruhr-Universit\"at Bochum, 44780 Bochum, Germany
             \\
             \email{fe@tp4.rub.de}
             \and Bay Area Environmental Research Institute, NASA Research Park, Moffett Field, CA, USA
             \and
             LESIA – Observatoire de Paris, Univ. PSL, CNRS, Sorbonne Univ., Univ. de Paris, 5 place Jules Janssen, F- 92190 Meudon, France
             \and
             University of Minnesota, Minneapolis, MN, USA
             \and
             European Space Agency (ESA), European Space Research and Technology Centre (ESTEC), Keplerlaan 1, 2201 AZ, Noordwijk, The Netherlands
             \and
         Institut f\"ur Experimentelle und Angewandte Physik, Universit\"at Kiel, 24118, Kiel, Germany \\
             }

   \date{}


  \abstract
   {}
   {We aim to constrain the acceleration, injection, and transport processes of flare-accelerated energetic electrons by comparing their characteristics at the Sun with those injected into interplanetary space. }
   {We have identified 17 energetic electron events well-observed with the SEPT instrument aboard STEREO which show a clear association with a hard X-ray (HXR) flare observed with the RHESSI spacecraft.
   We compare the spectral indices of the RHESSI HXR spectra with those of the interplanetary electrons. Because of the frequent double-power-law shape of the in situ electron spectra, we paid special attention to the choice of the spectral index used for comparison.}
   {
  The time difference between the electron onsets and the associated type III and microwave bursts suggests that the electron events are detected at 1 AU with apparent delays ranging from 9 to 41 minutes.
  While the parent solar activity is clearly impulsive, also showing a high correlation with extreme ultraviolet jets,
  most of the studied events occur in temporal coincidence with coronal mass ejections (CMEs).
  In spite of the observed onset delays and presence of CMEs in the low corona, we find a significant correlation of about 0.8 between the spectral indices of the HXR flare and the in situ electrons.
   The correlations increase if only events with significant anisotropy are considered. This suggests that transport effects can alter the injected spectra leading to a strongly reduced imprint of the flare acceleration.
   }
   {We conclude that interplanetary transport effects must be taken into account when inferring the initial acceleration of solar energetic electron events.
   Although our results suggest a clear imprint of flare acceleration for the analyzed event sample, a secondary acceleration might be present which could account for the observed delays. However, the limited and variable pitch-angle coverage of SEPT could also be the reason for the observed delays.
   }


   \maketitle
%

\section{Introduction}

In solar flares, energy that is stored in nonpotential coronal magnetic fields is released impulsively, presumably triggered by magnetic reconnection. In response, the solar atmosphere emits electromagnetic radiation over the whole wavelength range from radio to gamma rays \citep[e.g.,][]{Fletcher2011}. Analyses of this emission recorded by remote-sensing instruments have revealed key insights into the physics of solar flares. In particular, the observation of nonthermal bremsstrahlung in the hard X-ray (HXR) range has shown that electrons are efficiently accelerated in flares and carry a significant fraction of the energy released  \citep[cf.][]{Holman2011,Warmuth2020}. While nonthermal HXRs are primarily produced by electrons that precipitate into deeper layers of the solar atmosphere, electrons can also propagate upward through the corona and into interplanetary space.
 This is revealed by \mbox{type III} radio bursts, which are rapidly drifting structures observed in dynamic radio spectra \citep[cf.][]{White2011,Reid2014}. They are generated by escaping electron beams that excite Langmuir turbulence in the ambient plasma, which is subsequently converted to electromagnetic radiation.

Such interplanetary electron beams can be detected in situ by particle instruments on spacecraft.
However, one common feature of solar electron events has challenged our understanding of the parent acceleration process for decades, which is a frequently observed delay of 10-20 minutes between the occurrence of the solar counterpart, for example flare or radio type III burst, and the inferred injection time of the electrons based on their observed onset times at the spacecraft \citep[e.g.,][]{Haggerty2002, Kahler2007}. These delays were often interpreted as an indication of a different acceleration process, for instance a coronal or CME-driven shock \citep[e.g.,][]{Haggerty2003, Kahler2007}. However, also scenarios where flare-accelerated electrons are re-accelerated or energized by ongoing reconnection in the solar corona, possibly driven by the uplifting CME, are under discussion \citep{Maia2004}. Furthermore, magnetic trapping might also be involved in delayed onsets and for the acceleration of electrons to higher energies \citep[e.g.,][]{Dresing2018, Li2020}.

However, not all solar energetic electron events show such a delay and some case studies even reported different behavior at different energies, that is a prompt and a delayed component, within a single event \citep{Krucker1999, Wang2006, Li2020}. This suggests that some electron events detected in situ may indeed be of the same particle distribution as the HXR producing electrons, or at least still carry imprints of the flare acceleration. It is therefore tempting to compare the energy spectra of the HXR flare with the one of the in situ electrons.
 Depending on which approximation is used to invert the nonthermal HXR spectrum,  the expected relation of the photon spectral index $\gamma$ with the in situ electron spectral index $\delta$ is either $\delta = \gamma_{thick} +1$ for the thick-target model or $\delta = \gamma_{thin} - 1$ for the thin-target model \citep[cf.][]{Brown1971,Holman2011}.
 When analyzing 16 impulsive and non-delayed electron events observed by Wind/3DP \citep{Lin1995} and their HXR counterpart detected by the RHESSI spacecraft \citep[][]{Lin2002}, \citet[][]{Krucker2007} found a good linear correlation between the spectral indices of 0.83. However, the value pairs were neither consistent with the thin nor with the thick-target model but were all lying in between.
 Another sample of 15 events studied by \citet[][]{Krucker2007}, which was characterized by larger onset delays of $>8$ minutes, showed no clear correlation and a shift of the points toward the thin-target solution, which led \citet[][]{Petrosian2016} to conclude that the electrons of these events may have experienced a further acceleration after their initial energization in the flare. An additional argument for the close association of HXR-producing and escaping electrons has recently been provided by \cite{Xia2021} who reported on two events showing consistent energy cutoffs in the two populations.

The spectra of electrons observed in situ can often show spectral breaks or transitions that can relate to transport processes both near the original acceleration site and in interplanetary space \citep[e.g.,][]{Kontar2009, Strauss2020}. \cite{Dresing2020} studied the spectra of 781 electron events observed with the Solar Electron and Proton Telescope \citep[SEPT,][]{Muller-Mellin2007} aboard the two STEREO spacecraft and found double power-law shapes in 56\% of the events. In this paper, we use this electron event list as a starting point and investigate the relation between the energetic electron population precipitating onto the Sun as constrained by RHESSI, and the interplanetary electrons detected in situ with the SEPT instruments on board of the two STEREO spacecraft. We put a particular focus on the timing and spectral relation between both populations.

   \begin{figure}[t]
   \centering
  \includegraphics[ width=\hsize]{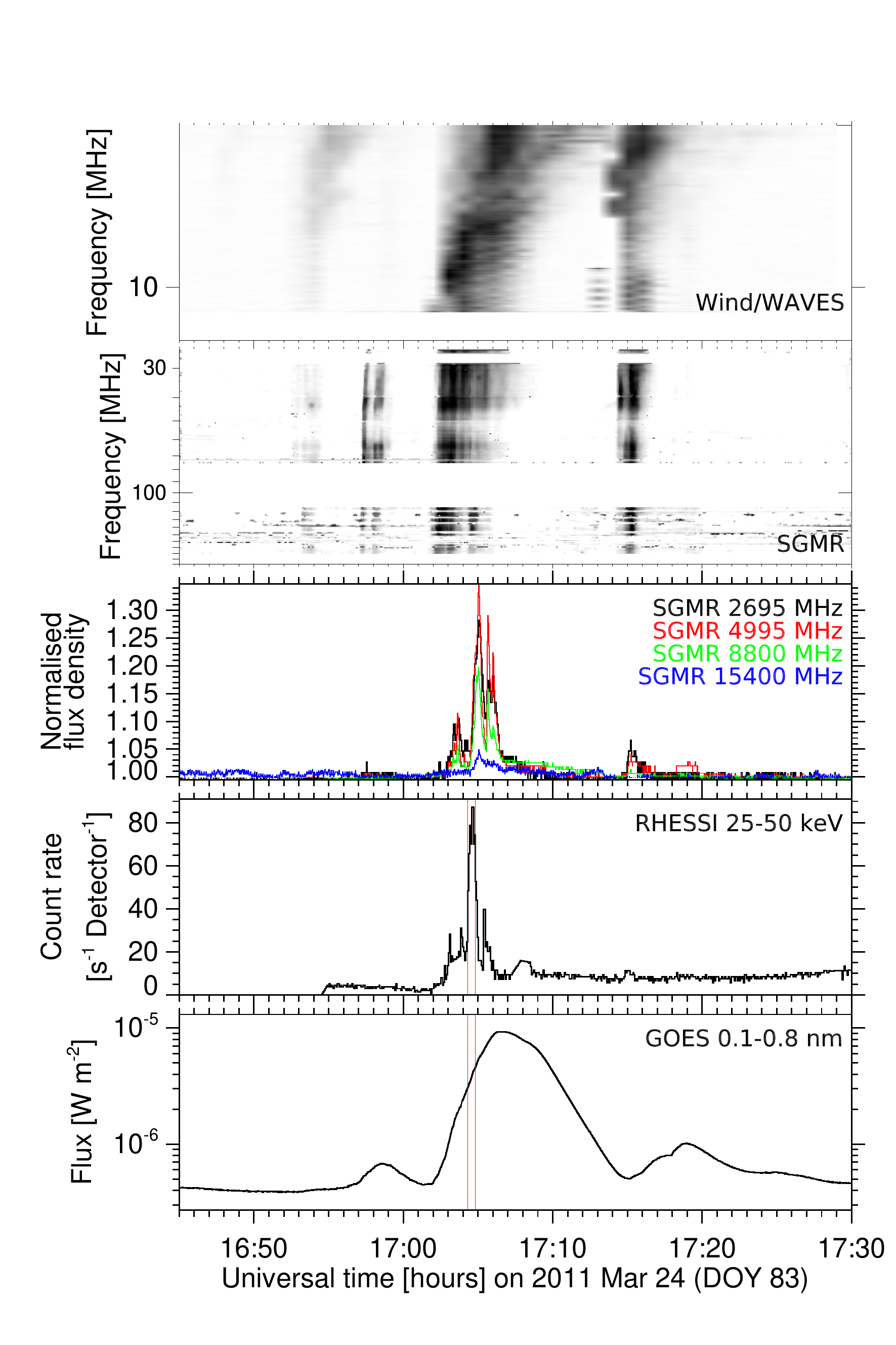}\\
      \caption{Time evolution of the X-ray and radio emission associated with the electron event on 2011 Mar 24. From bottom to top: time profiles of the GOES soft X-ray flux, of the RHESSI 25--50~keV count rates (corrected for instrumental effects), and of the radio flux at microwave frequencies, and dynamic spectra in the meter-wave and decameter-hectometer-wave radio range. The interval used for the computation of the RHESSI spectrum shown in Fig.~\ref{fig:hsi_spec} is indicated by the pair of red lines in the two bottom panels.}
         \label{Fig:radio_example}
   \end{figure}

   \begin{figure}[t]
   \centering
   \includegraphics[width=\hsize, trim=0cm 14.1cm 0cm 0cm, clip=true]{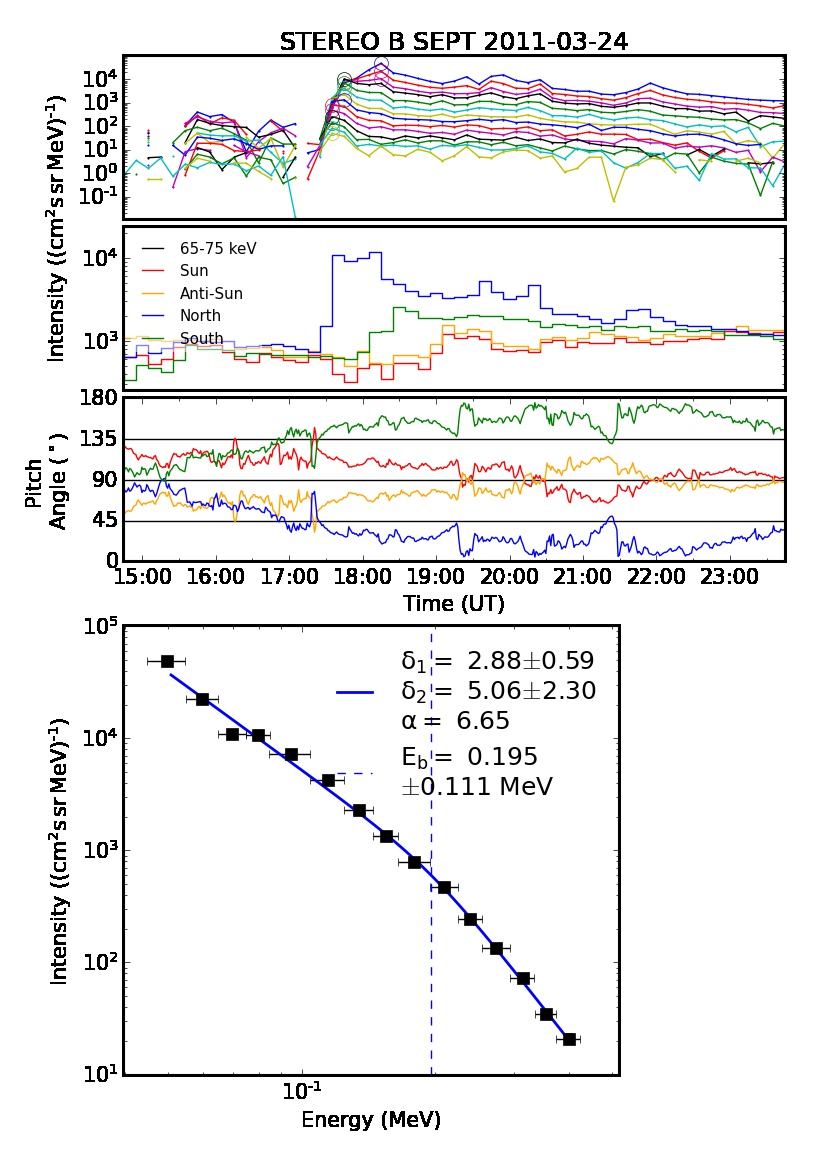}
      \caption{Solar energetic electron observations by SEPT aboard STEREO~B on 2011 Mar 24. The top panel shows the electron intensity in all available energy channels of one viewing direction with pre-event background subtraction, the second panel shows the 65-75 keV intensity in the four viewing directions of SEPT, and the third panel displays the corresponding pitch-angles based in the central pointing direction of the four telescopes.
              }
         \label{Fig:SEE_example}
   \end{figure}

\section{Observations and data selection}

Event selection started with a list of Solar Energetic Electron (SEE) events observed with the SEPT instruments on board of the two STEREO spacecraft compiled by \citet{Dresing2020} and available online\footnote{\url{http://www2.physik.uni-kiel.de/stereo/downloads/sept_electron_events.pdf}}. The version of the list used in this study covers the time from 2007 to 2018 and includes in total 925 SEE events, 557 at STEREO~A and 368 at STEREO~B. In the next step, the RHESSI flare list\footnote{\url{https://hesperia.gsfc.nasa.gov/rhessi3/data-access/rhessi-data/flare-list/index.html}} was searched for solar flares that had a start time within a one-hour window before the observed onset of the SEE event at the STEREO spacecraft. This yielded 64 SEE event candidates covered by SEPT as well as RHESSI. The low percentage of SEPT events recorded by RHESSI results from the fact that the STEREO spacecraft were magnetically not well-connected to the Earth-facing side of the Sun during the years of solar maximum, when the majority of interplanetary electron events was observed. This was especially the case for STEREO~A flying ahead of Earth leading its magnetic connection quickly to the backside of the Sun as seen from Earth. Therefore, all except the first two events in 2007 were detected by STEREO~B only.

In the next step, the association between the selected RHESSI flares and the SEPT events was ascertained by the geometric consideration of whether the flare was occurring at the  correct hemisphere so that a magnetic connection was at least remotely possible, followed by the inspection of radio spectrograms provided by the WAVES instrument on the Wind satellite \citep{Bougeret1995} and the SWAVES instruments on the two STEREO spacecraft  \citep{Bougeret2008}. \mbox{Type III} bursts showing close temporal association with the RHESSI flare were taken as strong evidence for an actual association. Some events had to be discarded due to issues with the RHESSI data, for example missing observations of the impulsive flare peak due to RHESSI nighttime. In this manner, we finally obtained 17 events, for which we then compared the spectral characteristics of the in situ electrons with the HXR photon spectra.
Two events on \mbox{2007 Jan 24} were observed by both STEREO spacecraft, which at that time were separated by only 0.5 degrees.
However, because the lowest energy channels were corrupted in early 2007 for STEREO~B data we only use the observations of STEREO~A for these two events in our correlation. Section \ref{app:multi-sc-events} in the appendix discusses the spectral variation between the closely spaced spacecraft for these two events and its implication for our study.

\begin{table*}
\caption{Event list including the basic parameters of the correlated flare and SEE events. For details, see main text.}
\label{tab:events}
         \centering
\begin{tabular}{ccccccccccc}
        \hline
        \hline
            \noalign{\smallskip}
   HXR & GOES & flare & & & & $E_\mathrm{b}$ & & $\Delta t$ & $L$ & $\Delta \Phi$, $\Delta \Theta$  \\
     peak time & class & location & $\gamma$ & $\delta_1$ & $\delta_2$ & [keV] & s/c & [min] &  [AU] & [deg] \\
    \hline
2007/01/24 00:31:24 & B5.1 & S05W61 & $3.2 \pm 0.09$ & $3.0 \pm 1.0$ & $3.8 \pm 0.6$ & 79 & A & $42\pm10$ &2.37 & 1, 1\\
2007/01/24 00:31:24 & B5.1 & S05W61 & $3.2 \pm 0.09$ & $2.7 \pm 1.0$ & $3.6 \pm 0.9$ & 107 & B & $43\pm1$ &2.49 & 1, 0\\
2007/01/24 05:16:09 & B6.8 & S05W64 & $4.1 \pm 0.26$ & $3.6 \pm 0.3$ & --            & -- & B & $37\pm1$ &2.09 & 3, 0\\
2007/01/24 05:16:09 & B6.8 & S05W64 & $4.1 \pm 0.26$ & $3.2 \pm 0.6$ & $3.9 \pm 1.3$ & 98 & A & $42\pm10$ &2.37 & 4, 1\\
2009/12/22 04:56:10 & C7.2 & S28W47 & $2.7 \pm 0.04$ & $2.0 \pm 0.1$ & $3.0 \pm 0.2$ & 122 & B & $31\pm1$ &1.75 & 35, 34\\
2010/02/08 03:12:24 & C6.2 & N22E00 & $3.4 \pm 0.10$ & $2.4 \pm 0.2$ & --            & --  & B & $34\pm1$ &1.92 & 7, 21\\
2010/11/12 03:53:41 & C1.0 & S21E02 & $3.7 \pm 0.06$ & $3.6 \pm 0.3$ &           --  &  --  & B & $37\pm1$ &2.04 & 29, 28\\
2010/11/12 08:01:54 & C1.5 & S23W01 & $3.9 \pm 0.08$ & $2.9 \pm 0.3$ & $4.5 \pm 0.5$ & 118 & B & $36\pm1$ &2.04 & 27, 30\\
2010/11/17 04:36:44 & B7.8 & S34E21 & $4.3 \pm 0.11$ & $1.6 \pm 1.7$ & $3.4 \pm 1.0$ & 69 & B & $47\pm1$ &2.6 & 10, 41\\
2011/03/24 17:04:34 & C9.1 & S15E41 & $4.1 \pm 0.05$ & $2.9 \pm 0.6$ & $5.1 \pm 2.3$ & 195 & B & $40\pm1$ &2.21 & 14, 14\\
2012/01/12 00:51:58 & C1.5 & N21E20 & $4.9 \pm 0.53$ & $4.4 \pm 0.4$ &  $6.4 \pm 2.0$ & 87   & B & $29\pm1$ &1.64 & 30, 14\\
2012/03/25 00:27:54 & C3.0 & N20E26 & $3.4 \pm 0.03$ & $2.7 \pm 0.4$ & $4.0 \pm 0.9$ & 110 & B & $50\pm1$ &2.83 & 23, 18\\
2012/04/16 00:26:14 & C1.8 & N13E89 & $4.9 \pm 0.10$ & $4.8 \pm 0.5$ &  --           &  --  & B & $39\pm1$ &2.21 & 10, 14\\
2012/05/07 03:21:58 & C2.7 & N13E67 & $4.4 \pm 0.03$ & $2.9 \pm 0.2$ & $4.9 \pm 0.6$ & 106 & B & $61\pm10$ &3.45 & 15, 17\\
2012/06/27 12:36:00 & C3.4 & N15E64 & $3.7 \pm 0.03$ & $2.4 \pm 0.3$ & $4.9 \pm 0.6$ & 90 & B & $29\pm1$ &1.64 & 1, 22\\
2012/06/28 02:12:24 & C2.6 & N17E56 & $4.3 \pm 0.09$ & $2.8 \pm 0.6$ & $3.7 \pm 0.4$ & 90 & B & $32\pm1$ &1.81 & 20, 24\\
2012/07/01 07:14:44 & C5.4 & N16E11 & $4.3 \pm 0.14$ & $3.6 \pm 0.3$ & $5.5 \pm 0.7$ & 104 & B & $31\pm1$ &1.7 & 54, 23\\
2014/03/19 16:26:14 & C3.3 & S12E81 & $5.9 \pm 0.20$ & $3.1 \pm 0.7$ & $4.0 \pm 0.7$ & 101 & B & $35\pm5$ &1.98 & 23, 18\\
2014/06/09 17:05:04 & C8.8 & S19E90 & $3.5 \pm 0.05$ & $2.6 \pm 0.5$ & $4.0 \pm 1.4$ & 90 & B & $43\pm1$ &2.43 & 28, 17 \\

    \hline
\end{tabular}
\end{table*}

\section{Analysis}\label{sec:analysis}
Figures \ref{Fig:radio_example} and \ref{Fig:SEE_example} show remote-sensing and in situ observations of an example event in the analyzed sample occurring on 24 Mar 2011. We present the analysis of both types of spectra in the following.

\begin{figure*}
\centering
\includegraphics[width=0.49\textwidth]{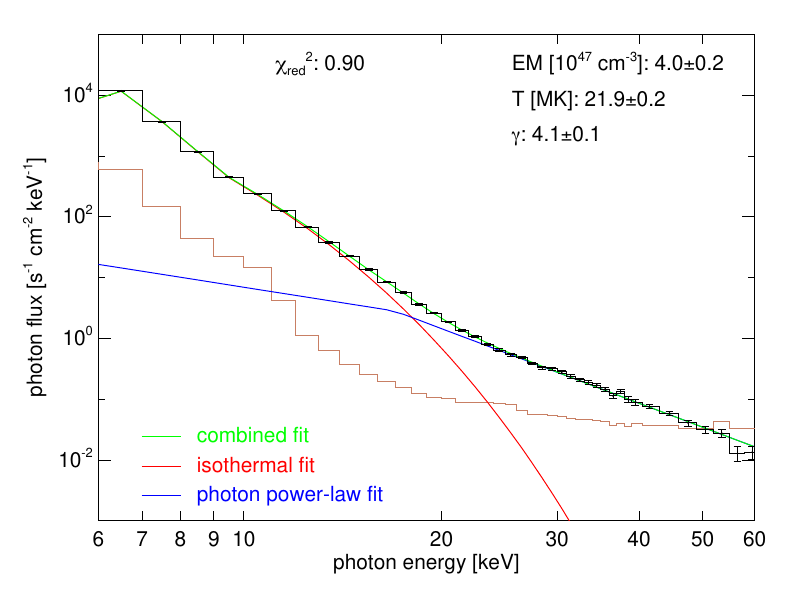}
\includegraphics[width=0.49\textwidth, trim=0cm 0.4cm 2.4cm 15.6cm, clip=true]{STEREO_B_example_event.jpg}
\caption{Left: Background-subtracted RHESSI photon flux spectrum (black) of the flare of 2011 Mar 24 fitted with an isothermal component (red) and a nonthermal photon power-law (blue). The spectrum was derived using all RHESSI detectors except for number 2 and 3. For comparison, the pre-event background (brown) is plotted as well. The fit results for emission measure EM, temperature $T$, and spectral index $\gamma$ are indicated. Right: Background-subtracted in situ electron spectrum of the 2011 Mar 24 event. The spectrum is constructed by the peak intensities of each energy channel (marked by circles in the top panel of Fig. \ref{Fig:SEE_example}). The line represents a broken-power law fit to the data.}
\label{fig:hsi_spec}
\end{figure*}

\subsection{X-ray and radio observations}
The time histories of X-ray and radio emission in Fig. \ref{Fig:radio_example} show a C9-class flare preceded and followed by distinct minor events. During the rise phase of the soft X-ray flux, bursts produced by nonthermal electrons are observed in hard X-rays and microwaves (second and third panels from bottom). The hard X-ray emission is bremsstrahlung produced by electrons of several tens of keV, the microwave emission is gyrosynchrotron radiation from electrons of several tens to hundreds of keV. The microwave data are from the Sagamore Hill (SGMR) Station (USA) of the Radio Solar Telescope Network (RSTN)\footnote{\url{https://www.ngdc.noaa.gov/stp/space-weather/solar-data/solar-features/solar-radio/rstn-1-second/}}. The two top panels display dynamic spectrograms in the (180-25) MHz range (SGMR\footnote{Data from \url{https://www.ngdc.noaa.gov/stp/space-weather/solar-data/solar-features/solar-radio/rstn-spectral/}}) and the (13.6-1) MHz range\footnote{\url{https://cdaweb.gsfc.nasa.gov/pub/data/wind/waves/}} \citep[Wind/WAVES;][]{Bougeret1995}. They show several groups of type III bursts, produced by electron beams at a few tens of keV propagating outward from the low corona along open magnetic field lines. The hard X-ray and microwave bursts in the low solar atmosphere occur at the times of well-identified and rather strong type III bursts. Besides at times of the main HXR and microwave emission there are tiny events, which accompany the other type III bursts. But the event clearly has one main episode of energetic electron acceleration in the low corona, lasting a few minutes, and the type III bursts demonstrate that simultaneously electrons escape along open field lines to the high corona and the interplanetary space.

The event is in many respects representative of our data set. Firstly, the electron events are accompanied by hard X-ray and microwave bursts and by metric-to-hectometric type III bursts. Secondly, the microwave bursts of 14 out of 17 events show time profiles and flux density spectra typical of gyrosynchrotron emission\footnote{We note that while it is in principle possible to constrain the electron spectrum from the optically thin gyrosynchrotron spectrum, a simple relationship between the two indices exists only in the extremely relativistic case, which generally does not apply to solar microwave bursts. In the case of the analyzed events, the microwave bursts are weak (a few tens of sfu at most), only a few percent of the quiet-Sun background. Although the nonthermal part can be identified, it is likely superposed on a thermal bremsstrahlung component. Under these circumstances, the microwave emission does not offer a valuable quantitative constraint of the electron spectral index.}, which suggests that electrons are accelerated to energies above 100 keV. In two other events the nonthermal microwave emission seems to be plasma emission, in one event only thermal bremsstrahlung is observed.
A third general feature is the gyrosynchrotron emission that lasts from 10~s to 5~min, and the nonthermal HXR emission lasting from 30~s to 2~min, in temporal coincidence with type III emission at m-$\lambda$.
Finally, in most events (13/17) several groups of type III bursts are observed, and only one is accompanied by a clear HXR and microwave burst.

\subsection{In situ observations of electrons}
The top panel of Fig.~\ref{Fig:SEE_example} displays the intensity measured by SEPT in all energy channels and the circles at the intensity maxima mark the peak intensities used to construct the background-subtracted peak intensity spectrum, which is shown in Fig. \ref{fig:hsi_spec} (right). The second panel from top shows the intensity of 65-75 keV electrons in all four viewing directions of SEPT with the corresponding pitch angles of the telescope center axes  plotted below. We note, that in case of anisotropic events (like shown in Fig.~\ref{Fig:SEE_example}), we always use the spectrum observed in the viewing direction measuring the highest intensity.
In case of isotropy, the Sun-facing telescope was used. A poor pitch-angle coverage due to non-nominal magnetic field configurations can seemingly reduce or even hide the real anisotropy of a particle beam. As in such cases the center of the particle beam with the highest electron intensities is not well observed, this can potentially also affect the determined spectrum and spectral shape (see also section~\ref{sec:correlations}).

\subsection{Analysis of photon and electron spectra}\label{sec:spectra_analysis}

For the associated HXR bursts, the peak time of the nonthermal emission was determined from RHESSI lightcurves by using the highest energy range in which a flare signature was observed. In most cases, this was in the range of 25-50~keV. In all cases, the used lightcurves showed a more impulsive behavior as seen in the 6-12~keV band, which is always dominated by thermal emission for medium to large flares. In all events, the duration of the nonthermal peak was around 30 sec. A RHESSI count spectrum was then obtained by integrating over a 30~sec time interval centered at this HXR peak time as marked by the vertical lines in Fig. \ref{Fig:radio_example}, using  a combination of all detectors that were well-functioning according to the RHESSI detector health database. Inspecting the background-subtracted spectra, we found the nonthermal part to be consistent with a single power-law in all events. The spectra were then fitted with the combination of an isothermal plasma component, a broken power-law with a fixed slope of -1.5 below the break that reproduces the nonthermal component \citep{Emslie2003}.
A photospheric albedo component was taken into account using the standard RHESSI OSPEX spectral analysis package\footnote{\url{https://hesperia.gsfc.nasa.gov/ssw/packages/spex/doc/ospex_explanation.htm} (based on \citet{Kontar2006}), assuming an isotropic electron distribution, which is more consistent with observations than a strongly beamed distribution \cite[cf.][]{Kontar-Brown2006}. We have additionally performed the spectral fitting without an albedo component and found only small differences of the spectral indices (smaller than 0.1) and conclude that the albedo component does not influence our results.}

The nonthermal emission is thus characterized by the spectral index $\gamma$ above the break. The upper energy limit of the fit range was determined by the counting statistics, which implied that several events could only be fitted up to 30~keV (the limit was 60~keV on average). Since photons of a given energy are mainly produced by electrons of about twice this energy, we do have an energy overlap with SEPT even in these cases. The resulting spectrum and corresponding fits for the example event on 24 Mar 2011 are shown in Fig.~\ref{fig:hsi_spec} (left). Additionally, we also performed thick-target fits to the spectra, which we use to determine the number of accelerated electrons (see Sect.~\ref{sec:elnum}).
Using the resulting spectral indices from these thick-target fits in our correlation analysis (not shown) does not change the correlation but as it involves extra model assumptions, we decided not to use these values. We furthermore refrained from using fits applying the thin-target model since we see no evidence for thin-target emission, as discussed in Section \ref{sec:correlations}.

\begin{figure*}[h]
\centering
\includegraphics[width=17cm]{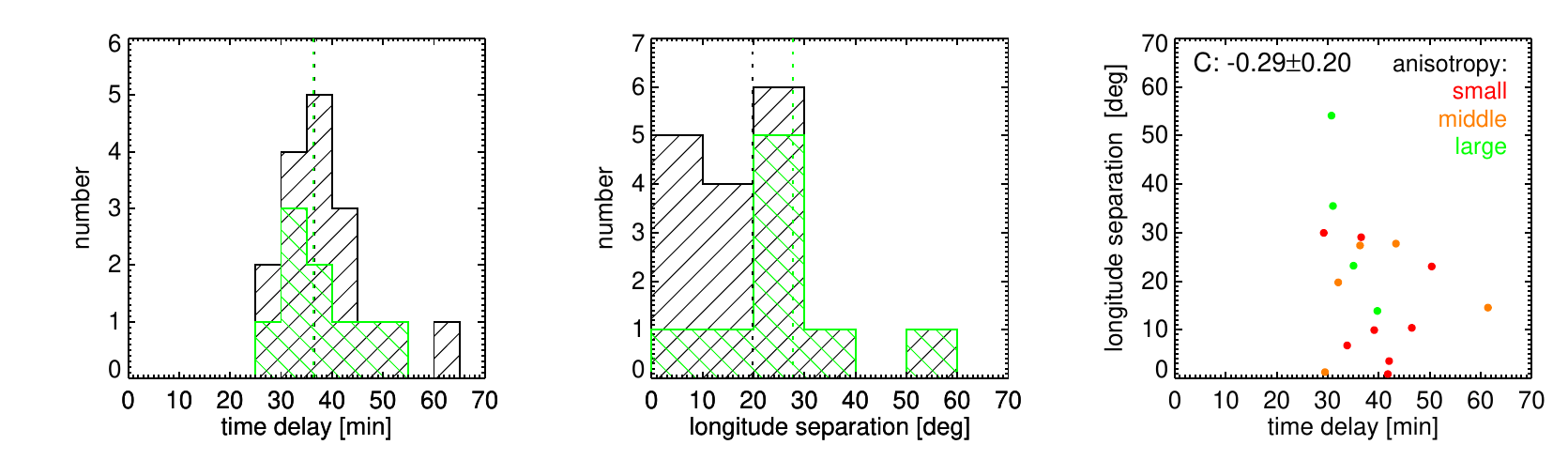}
\caption{\textit{Left}: Histogram of the time delay $\Delta t$ between the nonthermal HXR flare peak and the onset of the SEE event at STEREO. The black histogram refers to all events, while the green one shows the distribution for the events that are potentially associated with CMEs. The dotted lines indicate the median of the distributions. \textit{Middle}: Histogram of the longitude separation $\Delta \Phi$ between the flare location and the extrapolated footpoint of the magnetic field line connecting STEREO to the Sun. \textit{Right}: Correlation plot of the time delay $\Delta t$ and the longitude separation $\Delta \Phi$. The colors denote the degree of anisotropy of the in situ electrons. $C$ indicates the linear correlation coefficient.}
\label{fig:tdiff_longdiff}
\end{figure*}
The right-hand side shows the corresponding spectrum of the in situ electron event and a broken-power law fit.
For each event, the electron observations of SEPT were corrected from potential contamination due to ions or higher energy electrons as described in \cite{Dresing2020}. To determine the spectrum of the event we use the peak intensities observed individually in each available energy channel (marked by the circles in the top panel of Fig. \ref{Fig:SEE_example}) after a pre-event background subtraction has been applied.
We then fit each spectrum with single- and double-power law functions and chose the better fit based on the reduced chi square of the fits. The fits take into account the uncertainties of the peak intensities caused by counting statistics as well as the energy bin widths representing an uncertainty in energy. The uncertainties of the fit parameters were determined using 95\%-confidence intervals. For details see \citet{Dresing2020} and \citet{Strauss2020}.
We note that determining electron fluence spectra with STEREO/SEPT data is complicated due to the issue of ion contamination that is usually more dominant during the later phase of the events. Determining reliable fluence values and, especially, uncertainties, which are required by the fit, is difficult. We therefore decided not to present an analysis of the in situ fluence spectra here but comment on potential differences between the peak flux and the fluence spectra in Section \ref{sec:discussion_corr}.

Table~\ref{tab:events} shows the basic parameters for all 19 flare and SEE events, including the HXR peak time (as detected at the spacecraft), GOES class, flare location, HXR photon spectral index $\gamma$, in situ electron spectral indices $\delta_1$ and $\delta_2$\footnote{If $\delta_2$ and $E_b$ are missing, a single power law fit was used}, spectral break energy $E_\mathrm{b}$, which STEREO spacecraft detected the SEE event, the time delay $\Delta t$ between the nonthermal HXR flare peak (corrected for light travel time) and the onset of the SEE event at STEREO, propagation path length $L$ of 55-85 keV electrons corresponding to $\Delta t$ (see section \ref{sec:timing}), and the longitude and latitudinal separations $\Delta \Phi$ and $\Delta \Theta$ between the flare location and the ballistically extrapolated footpoint of the magnetic field line connecting the spacecraft to the Sun.

\subsection{Timing and magnetic connectivity}\label{sec:timing}

The onset times of the SEPT electron events were determined using the 3$\sigma$ method, which can only be considered as an upper limit for the real onset time.
An earlier onset could be masked by the background noise of the detector \citep{Laitinen2010}. Furthermore, especially in the case of more gradual intensity increases, which can also be caused by non-nominal magnetic field orientations resulting in a poor pitch-angle coverage at the SEPT instrument, the 3$\sigma$ method can yield onset times that are too late. Larger time averaging is often used to overcome these issues, and the time averaging applied for each event has been used as a measure for the uncertainty of the onset time \citep{Dresing2020}. This uncertainty has been propagated to $\Delta t$ in table~\ref{tab:events} in cases when higher averaging was used. Otherwise, the uncertainty represents the time resolution of the STEREO/SEPT data of 1 minute.
The time delays $\Delta t$ in our sample, which range from 29 to 61 minutes, with a median of $\Delta t$ = 37 minutes, are the delays between the HXR peak time at the Sun and the detection of the $55-85$ keV electron onset.
Assuming that the in situ measured electrons were injected at the time of the HXR peak, $\Delta t$ represents the propagation time of the first arriving particles.
A scatter-free propagation along a nominal Parker spiral would take about 20 minutes at these energies.
Comparing $\Delta t$ with this nominal and scatter-free propagation time we find that all events in our sample arrive delayed.

The events with medium to large anisotropy at their onset (see Table \ref{tab:additional_info}), where the spacecraft was likely well-connected to the electron source, have delays between 29 and 40 minutes. While these delays are among the shorter ones in our sample, they still exceed expectation for scatter-free propagation by 9-20 minutes.

Therefore, our event sample does not seems to contain any true prompt events as defined by \citet{Krucker2007} with a delay between the flare and the inferred injection time of the in situ measured electrons of $<8$ minutes. There can be several reasons for the observed delays. On the one hand, it could be due to instrumental effects leading to a delayed onset determination as described above. Furthermore, we note, that the pitch-angle coverage was not ideal in the majority of the events, which could cause apparent onset delays especially if the electron beam was very narrow during the early phase of the event. In this case, the delay would not represent a physical process related to the acceleration, injection, or transport of the particles. On the other hand, the actual propagation path of the electrons could be longer than the nominal Parker spiral, either due to large-scale distortions of the field or due to pitch-angle scattering. Field-line random walk caused by footpoint motion tied to the solar convection or by the turbulent solar wind evolution can cause significant lengthening of the actual path that particles need to follow along the magnetic field. Recent studies find that the effective field line length can be up to twice as long as the nominal Parker connection \citep[see, e.g.,][]{Laitinen2016,Laitinen2019,Chhiber2020}, which can cause equivalent delays in the observed particle onsets. Assuming that such effects or scattering cause the delays, i.e. the electrons were injected at the flare peak time, the column $L$ in the table provides the path lengths corresponding to $\Delta t$ for 55-85 keV electrons. Finally, an actually delayed injection of the electrons at the Sun can not be ruled out a priori. Different effects or a mixture of them being dominant for different events in our sample are, of course, also possible.

Figure~\ref{fig:tdiff_longdiff} shows histograms of $\Delta t$ and the longitudinal separations $\Delta \Phi$.
Despite the partially significant delays, the majority of the events is magnetically well-connected in longitude to STEREO, with $\Delta \Phi$ between 0.5 and 54 degrees (median: 15 degrees). There is no correlation between onset delay and longitude separation (see Fig.~\ref{fig:tdiff_longdiff} right). We also do not find any correlation with the latitudinal separation $\Delta \Theta$ (listed in Table~\ref{tab:events}). Furthermore, the degree of anisotropy (marked by the color in the right panel) does not correlate with the onset delays suggesting that interplanetary pitch-angle scattering is not the main reason for the observed delays.
The green histograms in Fig. \ref{fig:tdiff_longdiff} denote those events associated with CMEs that are located close to the flaring region and have heights below one solar radius at the time of the flare (see appendix \ref{appendix}). They may therefore potentially influence the injection and coronal propagation of the electrons.
While the events associated with these CMEs are not outstanding in other parameters of our analysis they seem to be among those events showing larger separation angles (middle panel of Fig. \ref{fig:tdiff_longdiff}) suggesting that the CME may be involved in enlarging the injection region.
The fact that many of the analyzed events are observed in different magnetic polarity sectors than that of the flaring active region at the Sun (see Table \ref{tab:additional_info}) could furthermore suggest an injection that covers a wider angular range even across the neutral line.
Such a scenario was also suggested by \citet{Kallenrode1993} who, however, reported that a current sheet crossing can also lead to a decrease in SEP intensities.
On the other hand, pitch-angle scattering in the IP medium may also be involved in filling both sides of the heliospheric current sheet with the electron population.
The majority of our events are also accompanied by coronal EUV jets (see appendix \ref{appendix}) but no particular connection with the time delay or separation angle was found.

\subsection{Spectral correlations}\label{sec:correlations}
\begin{figure*}
\centering
\includegraphics[width=8.5cm]{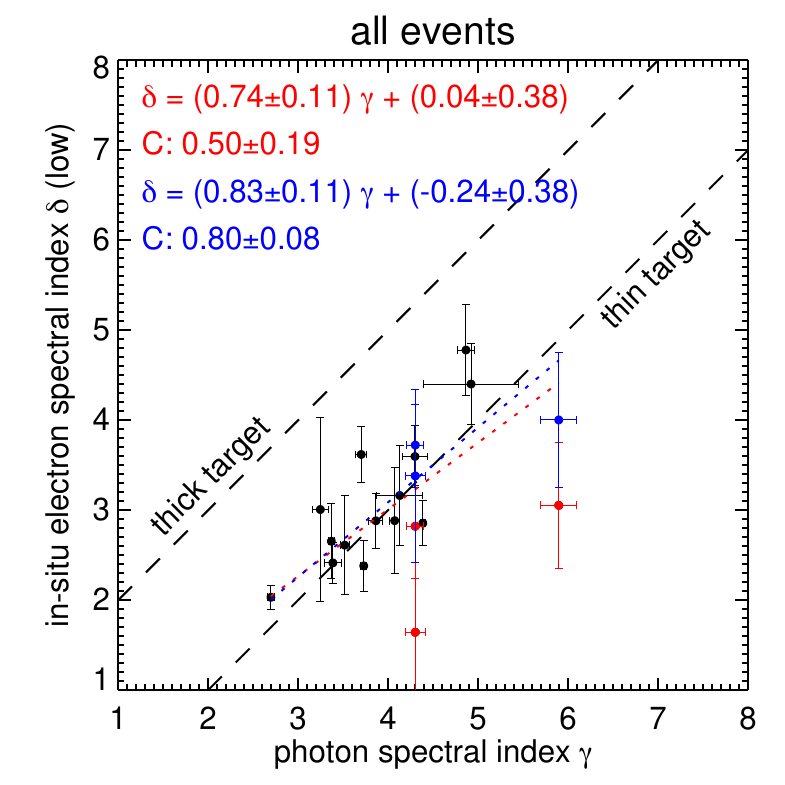}
\includegraphics[width=8.5cm]{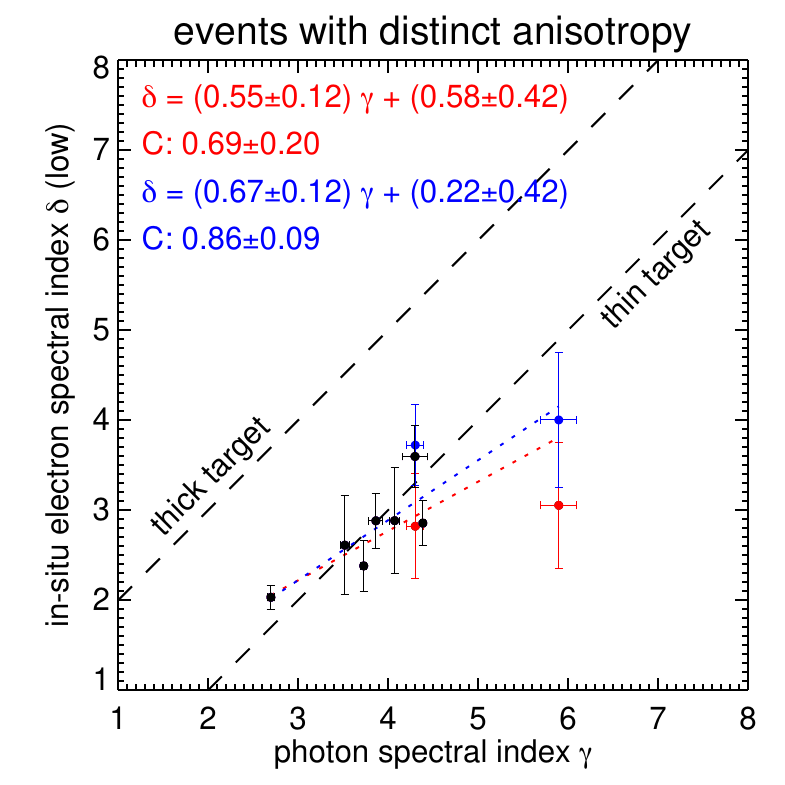}
\caption{Correlation plots of the HXR photon spectral index $\gamma$ observed by RHESSI and the electron spectral index $\delta$ obtained from STEREO/SEPT, plotted for all events (\textit{left}) and only the events with distinct anisotropy of the in situ electron flux (\textit{right}). In the case of events with broken power-laws, we use here the spectral index below the break, $\delta_1$. For three events, we consider either $\delta_1$ (red), or $\delta_2$ (blue). The dotted lines indicate linear fits to the data. The dashed lines indicate the relationships between $\gamma$ and $\delta$ expected for thick and thin-target bremsstrahlung, respectively. The fit parameters are indicated, as well as the linear correlation coefficients $C$.}
\label{fig:delta1_gamma}
\end{figure*}

\begin{figure*}
\centering
\includegraphics[width=8.5cm]{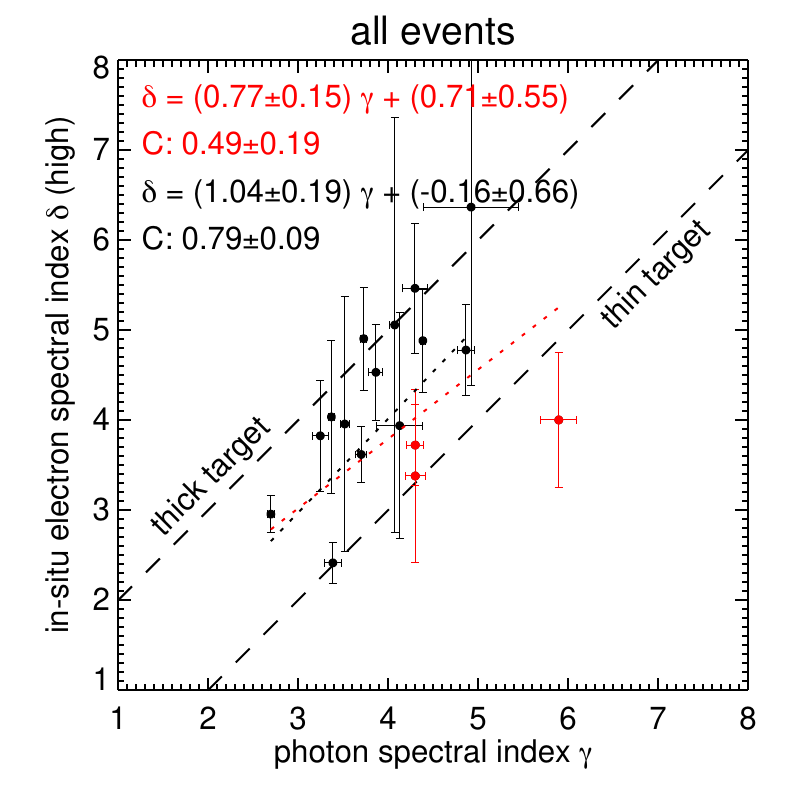}
\includegraphics[width=8.5cm]{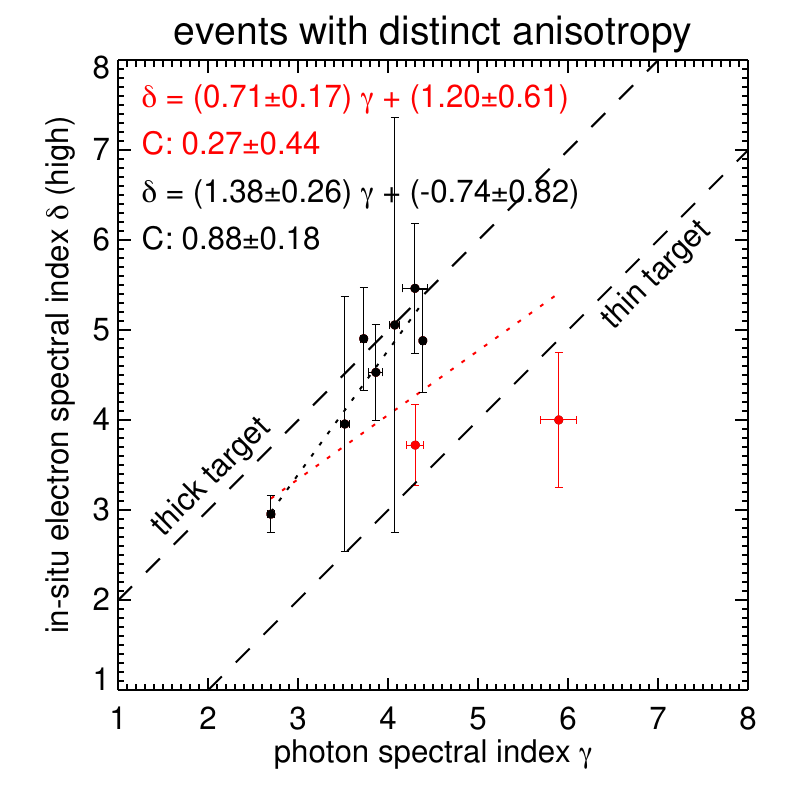}
\caption{As in Fig.~\ref{fig:delta1_gamma}, but showing the correlation of the HXR photon spectral index $\gamma$ with the electron spectral index $\delta_2$ (characterizing the high-energy part of the SEE spectrum) instead of $\delta_1$ in the cases of a broken power-law. We consider either all events including the three special cases (red), or we omit these events (black).}
\label{fig:delta2_gamma}
\end{figure*}

The results of correlating the spectral indices of the HXR photons and the in situ electron measurements are shown in Figures \ref{fig:delta1_gamma} and \ref{fig:delta2_gamma}.
As discussed above several energy-dependent effects may alter the spectrum observed in situ eventually leading to spectral breaks, which requires to make a choice between the two different spectral indices.
For the events in our sample this choice is often not straightforward as breaks are often found below or around 100 keV, which is in between the mean locations of the two different spectral breaks (60 keV and 120 keV) as reported for instance by, \citet{Krucker2007, Dresing2020}.
We therefore start by using always the lower ($\delta_1$) or the upper spectral index ($\delta_2$) as observed by SEPT in case of broken power-laws and compare these with the HXR spectral index $\gamma$.
The corresponding correlation plots are shown in figures \ref{fig:delta1_gamma} and \ref{fig:delta2_gamma}.
Note that in case of single power-law shapes the same in situ spectral index has been used in both plots. The panels on the left of the figures include all 17 events, while the ones on the right consider only the nine events that showed a distinct anisotropy (medium to large; see appendix~\ref{appendix} and last column of table~\ref{tab:additional_info}).

\begin{figure}
\centering
\includegraphics[width=8.5cm]{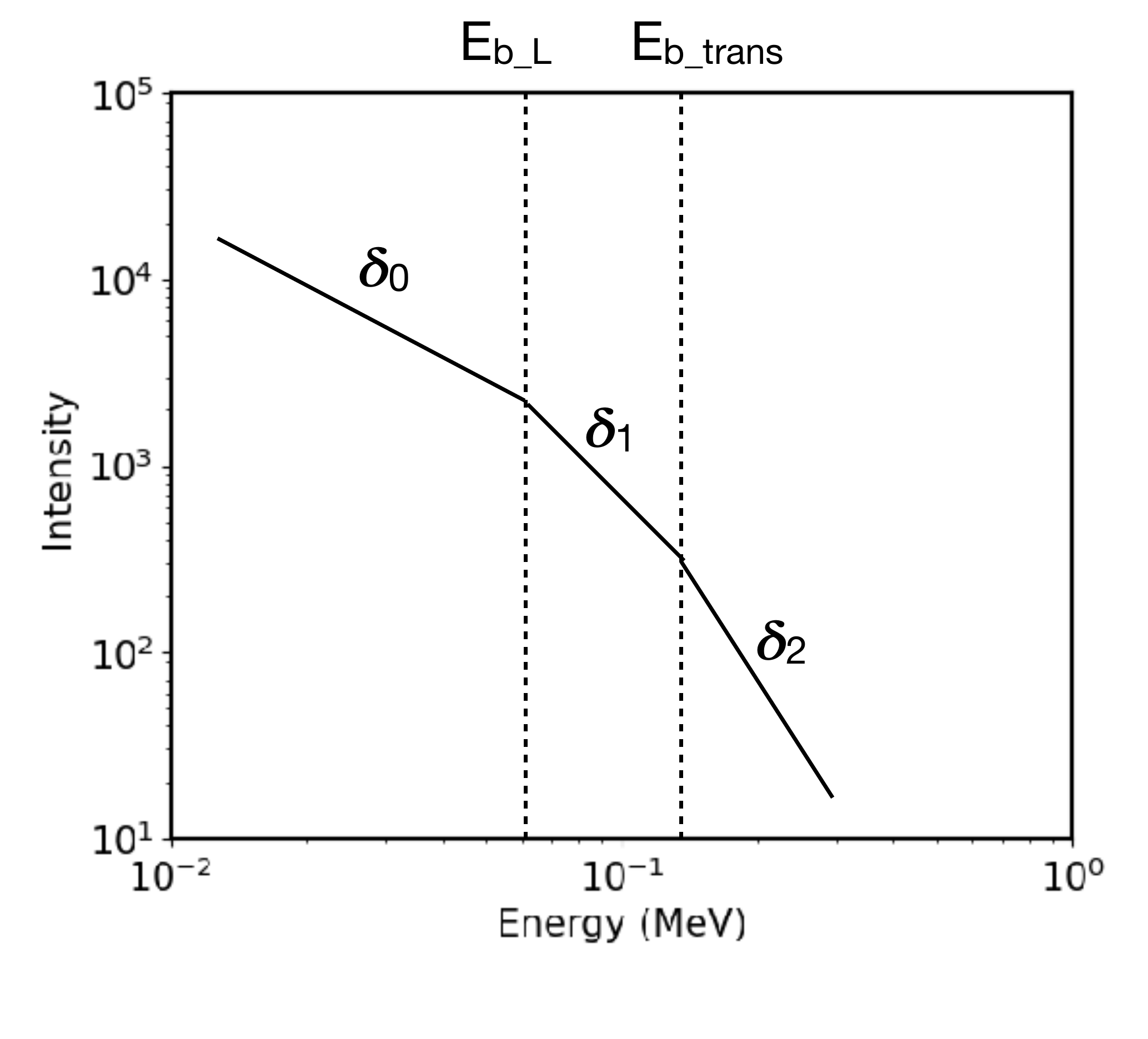}
\caption{Sketch illustrating a broken-power law spectrum with two breaks $\mbox{E}_{\mbox{b\_L}}$ and $\mbox{E}_{\mbox{b\_trans}}$ forming three spectral indices $\delta_0$, $\delta_1$ and $\delta_2$.}
\label{fig:spec_sketch}
\end{figure}

Three events appear to behave differently as their spectral indices $\delta_1$ and $\delta_2$ lie systematically below the rest of the distributions (marked by colored points in figures~\ref{fig:delta1_gamma} and \ref{fig:delta2_gamma}).
We suspect that these three cases represent events showing a spectral break caused by Langmuir-wave generation (expected to occur at lower energies than the break due to pitch-angle scattering) while the rest of the broken-power-law events corresponds to breaks caused by pitch-angle scattering (see appendix \ref{app:three-special-events} for more discussion on these events).
Fig.~\ref{fig:spec_sketch} helps to illustrate this. It shows a broken power-law spectrum containing two breaks and therefore three different spectral indices $\delta_0$, $\delta_1$ and $\delta_2$.
The first break, marked by $\mbox{E}_{\mbox{b\_L}}$  is assumed to correspond to Langmuir-wave generation and the second break  ( $\mbox{E}_{\mbox{b\_trans}}$)  to pitch-angle scattering during interplanetary transport, respectively.
According to the description above, the three special events would cover the spectral part of $\delta_0$ and $\delta_1$ while the other events correspond to $\delta_1$ and $\delta_2$ in this sketch.
To correctly handle the spectral indices of these three events we therefore have to treat their $\delta_2$ values as $\delta_1$ values. Their $\delta_1$ would correspond to $\delta_0$ in the sketch (not covered by the other events) and the $\delta_2$ range of the other events is not covered by these three events.
We therefore have to use the blue points instead of red points in Fig.~\ref{fig:delta1_gamma} and remove the $\delta_2$ values (red points) from the correlation in Fig.~\ref{fig:delta2_gamma}.

Figure~\ref{fig:delta1_gamma} shows that $\gamma$ and $\delta_1$ are correlated. For the three events discussed above, both spectral indices $\delta_1$ (red, suspected wrong values) and $\delta_2$ (blue, suspected correct values) are included. The differently colored legends represent the Pearson correlation coefficient and linear fit results using either $\delta_1$ (red) or $\delta_2$ (blue) of these three events.
Similarly, for Fig.~\ref{fig:delta2_gamma} the correlation in red corresponds to using all $\delta_2$ values in the sample, and the one in black excludes the three red points as discussed above assuming that these events do not provide a corresponding $\delta_2$ observation.
The correlations in both plots improve significantly when treating the three events as discussed. Although the correlations with the lower and higher in situ spectral indices are very similar, the implications for the relation between the photon and in situ spectral indices are very different: The value pairs in Fig.~\ref{fig:delta1_gamma} clearly align along the thin-target approximation, while the points in  Fig.~\ref{fig:delta2_gamma} shift to the range between thick and thin-target lines with some points even lying above the thick target model.

Taking the correlation of $\gamma$ and $\delta_1$ at face value, one could conclude that the nonthermal photon spectrum is actually generated by thin-target emission, and that indeed both the remote-sensing and the in situ observations detect the same particle population. However, thin-target emission is generally believed to dominate nonthermal flare spectra only in those cases where the HXR footpoints are occulted by the solar limb \citep[cf.][]{Krucker2008rev}. The footpoints are the locations where accelerated electrons are stopped collisionally in the denser chromosphere, and consequently they emit thick-target radiation, which will usually dominate any thin-target contribution from the corona. We have performed HXR imaging with RHESSI for all flares in order to ascertain whether HXR footpoints are present or could be potentially obscured. In 7 flares, footpoint pairs could be unambiguously detected, while in another 7 events, only possible indications for footpoints could be found due to the limited image fidelity caused by the small number of nonthermal counts. In the rest of the events, nonthermal and thermal sources were found to be contiguous. Only one flare was located so close to the limb that an occultation of the footpoints is possible. We find no systematic differences in the spectral correlations for the events with clearly visible footpoints as compared to the other flares. We thus conclude that the HXR spectrum is actually dominated by thick-target emission.

Table \ref{tab:correlations} summarizes the determined correlation coefficients between the photon spectral index $\gamma$ and the two electron spectral indices  $\delta_1$ and $\delta_2$, also including further subsamples based on a rough anisotropy classification of the events (see appendix~\ref{appendix} and last column of table~\ref{tab:additional_info}).
The correlations clearly improve if only anisotropic events are taken into account, however the number of events with large anisotropies is unfortunately very low.
We also investigated the quality of the correlations with respect to the onset delay, the longitudinal, and latitudinal separation angles between flare and spacecraft magnetic footpoint at the Sun as well as the presence of CMEs or EUV jets but did not find any dependence on these parameters.
\begin{table}
\caption{Pearson correlation coefficients $C$ between photon spectral index $\gamma$ and in situ electron spectral index $\delta$. Also given are the uncertainties on $C$ based on a bootstrapping approach.}
\label{tab:correlations}
         \centering
\begin{tabular}{lccc}
        \hline
        \hline
            \noalign{\smallskip}
  & all & medium \& large & large \\
  & events & anisotropy & anisotropy \\
    \hline
no. of events  & 17 & 9 & 4\\
\hline
$\gamma$ vs. $\delta_1$ & $0.50 \pm 0.19$ & $0.69 \pm 0.20$ & $0.61 \pm 0.55$  \\
$\gamma$ vs. $\delta_2$ & $0.49 \pm 0.19$ & $0.27 \pm 0.44$ & $0.28 \pm 0.77$ \\
$\gamma$ vs. $\delta_1$\tablefootmark{1} & $0.80 \pm 0.08$ & $0.86 \pm 0.09$ & $0.96 \pm 0.06$ \\
$\gamma$ vs. $\delta_2$\tablefootmark{2} & $0.79 \pm 0.09$ & $0.88 \pm 0.18$ & $1.0 \pm 10^{-4}$ \\
    \hline
    \end{tabular}
    \tablefoot{
   \tablefoottext{1}
   In situ electron spectral index $\delta_2$ has been used for the three special events (see text).
   \tablefoottext{2} In situ electron spectral index $\delta_2$ has been excluded for the three special events (see text).
   }
\end{table}

\begin{figure*}
\centering
\includegraphics[width=8.5cm]{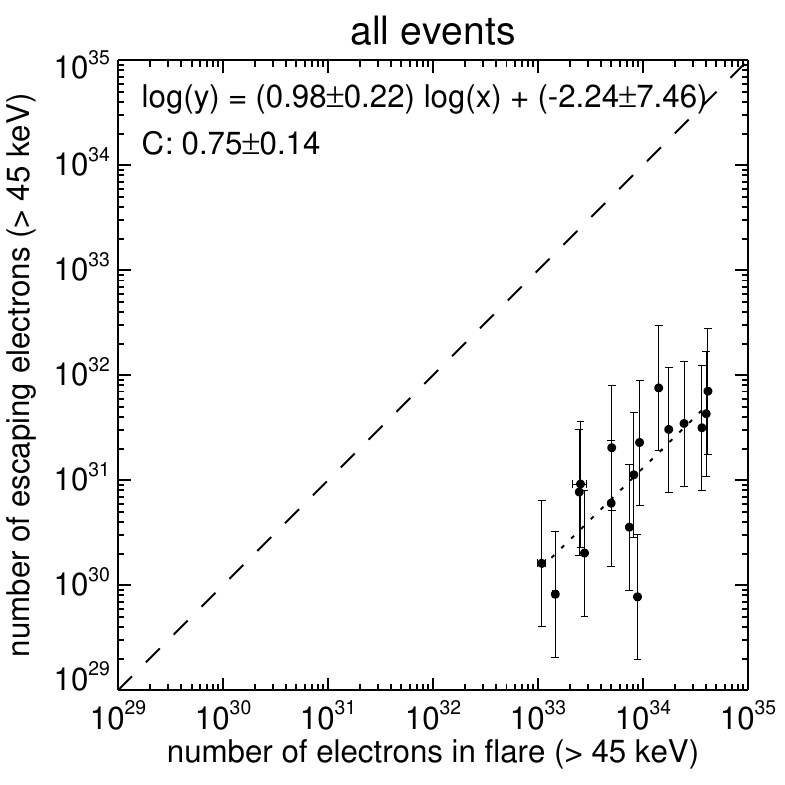}
\includegraphics[width=8.5cm]{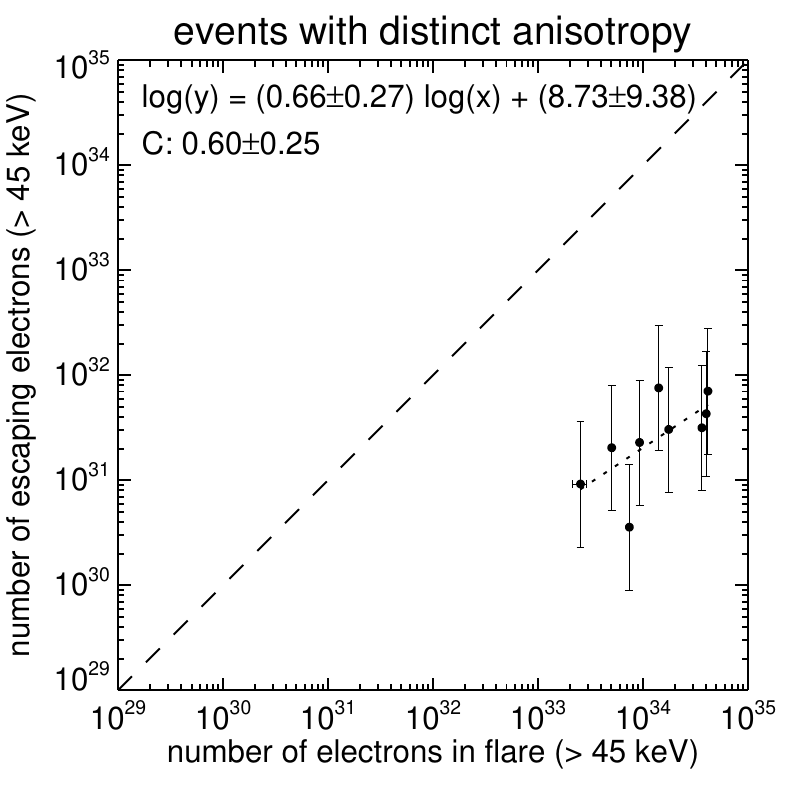}
\caption{Correlation plots of the number of nonthermal electrons above 45~keV in the flares derived from thick-target fits of HXR spectra and the number of electrons escaping into interplanetary space, plotted for all events (\textit{left}) and only the events with distinct anisotropy of the in situ electron flux (\textit{right}).}
\label{fig:enum}
\end{figure*}

\subsection{Total number of escaping electrons}
\label{sec:elnum}
Figure~\ref{fig:enum} shows the total number of escaping electrons inferred from the in situ observations ($>45$ keV and $<425$ keV) as a function of the number of electrons $>45$ keV accelerated in the flare assuming thick-target emission and integrating the derived peak flux over 30 seconds.
The number of in situ electrons has been determined by integrating the fluence spectrum of the events observed in the telescope showing the highest intensity increase, which was also used to construct the peak intensity spectrum. The fluence is the integrated flux over the duration of the event, with the duration being defined by the onset of the event and an individual end time. This end time was determined for each energy channel individually by the time when the flux decreased to $1/e$ of the peak flux. In the same manner as for the peak intensity spectra, analyzed in this work, the pre-event background was subtracted from the flux before calculating the fluence and the contamination correction was applied (c.f. Section~\ref{sec:spectra_analysis}) assuming a constant contamination over the duration of the event.

Following \cite{Krucker2007} we assume an electron beam that is emitted into a cone with a width of $30^{\circ}$ to derive the total number of in situ electrons, and we use cones of $15^{\circ}$ and $60^{\circ}$ to estimate the uncertainties shown in Fig.~\ref{fig:enum}. Propagating this cone along the nominal Parker spiral to a distance of 1 AU results in a spherical cap over which the electrons are spread at the spacecraft position. To infer the total number of electrons of the event, the number of electrons detected in the small SEPT instrument are therefore scaled to the area of this spherical surface.
We note, that this approach assumes a constant energetic electron density over the cone angle, which is only a first-order approximation. Secondly, the sectored measurements by SEPT show that the observed particle distribution at the spacecraft is much more isotropic compared to the assumed cone, which is expected in case of non-negligible particle scattering during interplanetary transport. However, without proper transport simulations and multi-spacecraft observations a more accurate approach demanding fewer assumptions to determine the total number of in situ electrons is not feasible.

The left hand plot of Fig.~\ref{fig:enum} shows all events in our sample, while the right hand figure shows only events with significant anisotropy. The legends provide the results of linear fits and the Pearson correlation coefficients. A reasonable correlation of 0.75 is found for all events, and a slightly lower correlation of 0.6 for the anisotropic events. Although one would rather expect a larger correlation for anisotropic events, i.e. in the case of less pitch-angle scattering, an effect that can reduce the determined number of escaping electrons, the anisotropic sample does indeed contain on average higher numbers of escaping in situ electrons. In contrast to results by \cite{James2017}, who used data by ACE/EPAM \citep{Gold1998} and determined a fraction of 6 to 148\% of escaping electrons compared to the HXR-producing electrons our results confirm the findings by \cite{Krucker2007}. We find very small fractions of escaping electrons with mean ratios of only 0.18\% for all events and 0.24\% for the anisotropic events. For comparison, \cite{Krucker2007} found a ratio of 0.2\%.

\section{Discussion}

\subsection{Modification of electron spectra between the acceleration region and the spacecraft}
\label{sec:transport_effects_on_spec}
Transport processes in the flare and in interplanetary space can alter the initial spectral distribution of nonthermal electrons. One such process that causes energy loss is due to Coulomb collisions \citep[cf.][]{Brown1971} when energetic electrons propagate through a background plasma.
For the escaping electrons, this effect is negligible in interplanetary space, but it can play a role near the acceleration site. \cite{Reid2013} concluded that coulomb collisions are negligible as an energy-loss process above an energy near 40 keV, which is in the relevant range for our analysis. However, because collisional energy loss will lead to a progressive flattening of the electron flux spectrum toward lower energies, it could thus potentially account for the fact that the in situ spectra are harder than the ones inferred from HXR observations (see Fig. \ref{fig:delta1_gamma}). We have, therefore, considered a model of the ambient electron density consisting of a 2.5-fold Newkirk model \citep{Newkirk1961} in the lower corona, which then transitions to the Mann model \citep{Mann1999} that is more appropriate for the upper corona and IP space. We inferred the electron density in the corona above the acceleration region from the starting frequency of the type III bursts, as observed by the worldwide e-Callisto network (\url{www.e-callisto.org}). We found the median start frequency near 330 MHz. Under the usual assumption of harmonic plasma emission this corresponds to an ambient electron density of $3.4 \cdot 10^8$ cm$^{-3}$ and an acceleration region below 0.1 R$_{\odot}$ above the photosphere. Taking this value, integrating from this height to 1 AU, and computing the effect of the spectral flattening according to Brown (1971), we find that Coulomb collisions have indeed a negligible effect on the spectral index in the energy range we have considered here, and thus cannot account for the spectral differences between precipitating and escaping electrons.

While Coulomb collisions may be neglected in our analysis, other processes such as the generation of Langmuir waves or pitch-angle scattering may cause spectral changes both in the flare and in the interplanetary medium (as discussed in Section \ref{sec:correlations}).  \cite{Kontar2014} showed for example that strong pitch-angle scattering in coronal loops can potentially cause a flattening of the HXR spectrum and lead to spectral breaks.
However, several previous studies have found significant correlations between the HXR and in situ electron spectral index \citep[e.g.,][]{Kallenrode1987, Droege1996, Krucker2007} and it was suggested that 1) both spectra belong to the same accelerated population and 2) that transport effects do only play a minor role in altering the spectra. Many of such past studies, which analyzed solar energetic electron spectra have used instruments, which covered only energies $\gtrsim100$ keV and which lacked a fine energy resolution in the lower energy part that is needed to resolve spectral breaks caused by interplanetary transport processes, i.e. the generation of Langmuir turbulence and pitch-angle scattering, as discussed in this manuscript. Furthermore, both of the mentioned transport processes are dominant at lower or near-relativistic energies \citep{Droege2003, Agueda2014} making it difficult to detect these with instrumentation covering mainly $\gtrsim100$ keV.
Instruments such as the Wind/3DP and STEREO/SEPT mainly measure near-relativistic electrons where these spectral breaks occur.
However, because of the slightly different energy coverage and resolution of the Wind/3DP and STEREO/SEPT instruments they are likely sensitive to different spectral breaks with 3DP usually detecting the break caused by Langmuir-wave generation and SEPT the one by pitch-angle scattering \citep[see][]{Krucker2009, Kontar2009, Dresing2020, Strauss2020} so that the spectral index above the break $\delta_2$ detected by 3DP is likely covered by the spectral index below the break $\delta_1$ as seen in SEPT data. However, an unambiguous identification of the type of the spectral break in a single event can be difficult because of the overlapping energy ranges of these effects. Three events in our sample showed a systematic shift with respect to the spectral indices of the rest of the events suggesting that these were indeed events covering the Langmuir wave-related spectral break different to the other events. Therefore, these events have been re-categorized accordingly (see section~\ref{sec:correlations} and figures~\ref{fig:delta1_gamma} and \ref{fig:delta2_gamma}) leading to significantly improved correlations.

\subsection{The spectral correlation of HXR-producing and in situ electrons}\label{sec:discussion_corr}
We find correlations between the HXR spectral index and the one of the in situ electron spectrum of about 0.8 for both sets of value pairs either using the spectral index below or above the break (see Table \ref{tab:correlations}), however when using $\delta_1$ the points align along the thin-target solution (Fig. \ref{fig:delta1_gamma}), while they are shifted more toward the thick-target line (lying between thick and thin-target lines) when using $\delta_2$ (Fig. \ref{fig:delta2_gamma}).
A preliminary analysis of the corresponding SEPT fluence spectra showed overall no change for the $\delta_1$ values, but the $\delta_2$ values showed a trend of shifting toward softer spectral indices causing the data points to align more along the thick-target solution or lying even above it in the correlation plot (not shown). However, due to the issue of ion contamination these fluence spectra suffer large uncertainties and are less reliable.

We assume that our value pairs using $\delta_1$ should be compared to the values shown by \cite{Krucker2007} who compared HXR spectra observed by RHESSI with Wind/3DP electron spectra above the break ($\delta_2$). They found a similar correlation of about 0.8, however, their value pairs were lying between the thick and thin target solutions.
The alignment of our value pairs using $\delta_1$ along the thin-target model is not expected since imaging of nonthermal footpoints shows that in most events thick-target emission has to be dominant. After excluding also the role of energy loss due to Coulomb collisions (see Section \ref{sec:transport_effects_on_spec}) we therefore suspect that another systematic effect causes this shift.
\cite{Krucker2007} found only a good correlation for non-delayed events suggesting that the delays, frequently observed during solar energetic electron events \citep[e.g.,][]{Haggerty2002, Kahler2007}, may be caused by another or an additional acceleration process.
This is different for the events studied in this work: all of our events show delays of at least 9 minutes with respect to the flare. However, as discussed in Section \ref{sec:timing} instrumental effects and especially nonideal pitch-angle coverage could often be the cause for apparent delays.

Although \cite{Krucker2007} found only a low correlation of 0.43 for their delayed event sample, those value pairs did also show a shift toward the thin-target line, i.e. toward harder in situ electron spectra like our $\gamma-\delta_1$-value pairs.
\cite{Petrosian2016} suggested that a further acceleration process acting on the flare-accelerated electron distribution could be the reason for this shift. However, given the still significant correlation coefficient for our value pairs we note that such a further acceleration should either only be of minor importance or scale with the flare itself, so that only a systematic shift of the spectral index is caused, and the overall correlation is preserved.
We note that \citet{Krucker2007} also suggested the possibility of further acceleration in the flare due to trapping in closed and shrinking field lines. This would, however, only affect the downward moving electrons, which produce the HXR spectrum.

\subsection{Anisotropic electron events and the number of escaping electrons}
We find a clear improvement of our correlations when only taking into account anisotropic events suggesting that pitch-angle scattering can lead to a vanishing imprint of the acceleration. We note that the lack of anisotropy can either be caused by strong scattering during interplanetary transport but also due to poor pitch-angle coverage of SEPT caused by non-nominal magnetic field configurations. See table~\ref{tab:additional_info}, which lists the strength of the anisotropy for each event and marks if poor pitch-angle coverage was present.
Consequently, even in the case of low scattering conditions, the limitations of the measurement can lead to vanishing correlations, i.e. to a change of the observed spectral indices.
We note, that Wind/3DP does not suffer from such a limitation because of its unique directional observations covering the complete 4$\pi$ space, which is for example not the case for the ACE/EPAM instrument \citep{Gold1998}. Although EPAM/LEMS provides eight different sectors determined through the spacecraft's spin, there exist magnetic field directions that are perpendicular to the measurement plane of EPAM/LEMS reducing the pitch-angle coverage to only one point in pitch-angle space for all sectors.

The anisotropy is also expected to influence the correlation between the number of escaping electrons with the number of flare electrons as shown in Fig.~\ref{fig:enum} because larger anisotropies imply weaker interplanetary scattering conditions and therefore less dispersal of the injected population. However, we do not find an improvement of the correlations when only taking into account the anisotropic events. But poor pitch-angle coverage, which can lead to the underestimation of the anisotropy as well as to an underestimation of the total number of electrons, may again affect the results. Furthermore, the strong assumptions, which have to be made to determine the number of escaping electrons likely play an even larger role for the weak correlation. These assumptions are i) the same size of the cone ($30^{\circ}$) filled with electrons for each event, which is very likely not true due to varying injection cone sizes, i.e. the angular width of open magnetic field lines.
ii) A constant distribution of electrons over the $30^{\circ}$ cone, which is only a first-order approximation because of the observed presence of weak or missing anisotropies depicting the presence of interplanetary scattering. Furthermore, an angular-dependent injection function is possible. iii) A nominal Parker field with a fixed Parker spiral length for all events, not taking into account large-scale structures or field line meandering, which will further introduce event to event variations leading to a vanishing correlation.
Nevertheless, our results confirm the findings by \citet{Krucker2007} that only a very small fraction ($\sim 0.2$\%) of accelerated electrons are finally injected into interplanetary space when compared to the number of the HXR producing electrons.

\subsection{The timing of electron release in the corona}
A simple picture of the relationship between electrons detected at 1 AU and radio or X-ray emitting electrons in the solar atmosphere is a common short acceleration and the release into magnetic structures in the corona and onto open field lines to the heliosphere. This picture is supported by a number of detailed timing studies of the start times of impulsive electron events \citep[see the discussion of the 2000/05/01 case in][]{Kle-21b}.
But in other events the first electrons were reported to arrive later at the spacecraft than expected. Instruments with limited pitch-angle coverage, such as the STEREO detectors, might just miss the first arriving electrons, creating an apparent delay.
Linhua Wang and coworkers \citep{Wang2006,WnL:al-16} exploited the excellent pitch-angle coverage of the Wind/3DP instrument in systematic analyses of the onset time and the duration of release of electrons across the energy range between a few keV and some tens or hundreds of keV. They found that electrons detected with energies below 20 keV at the spacecraft were released since earlier times and over longer durations than the electrons above about 30 keV.
\cite{Kontar2009} ascribe such inferred earlier releases to the energy loss of the Langmuir-wave generating (low-energy) electrons. As an alternative, or an addition,  our observations suggest a scenario where electron beams are accelerated to different energies in several successive episodes: we find that the solar counterparts of the electron events are in general groups of type III bursts, discernible at frequencies above $\sim$1 MHz. As illustrated in Fig. \ref{Fig:radio_example}, the hard X-ray and microwave emissions, which are produced by electrons at tens to hundreds of keV, accompany one of these groups of type III bursts, but other groups occur before and afterward. Such additional type III bursts without prominent microwave counterpart are observed in 13/17 of our events. This is also shown  by the timing of the microwave bursts and the DH type III bursts (Table \ref{tab:additional_info}, cols. 2 and 4). Since type III bursts at 1 AU are produced by electrons with energies below 20 keV \citep[e.g.,][]{Erg:al-98}, this timing implies that repeated episodes of impulsive electron acceleration to relatively low energies (a few tens of keV) occur in the corona, and that one such episode is also accompanied by the acceleration of near-relativistic electrons. The low-energy electrons are hence  accelerated over longer times than those of higher energies that are seen through their hard X-ray and microwave emission, consistent with Wang and coworkers. But the higher time resolution offered by the radio and HXR analysis shows that the observations cannot be ascribed to a single, time-extended acceleration episode of low-energy electrons, and a different, shorter and delayed, episode at the higher energies. The electron fluxes at 1 AU rather result from multiple injections of electrons in the corona. Each release produces electrons up to a few tens of keV that are able to emit type III bursts, but the near-relativistic electrons are only accelerated during part of these acceleration episodes. This particular interval actually consists of several successive events, too, as seen by the multiple microwave peaks. \cite{Vla:Rao-95} argued indeed that apparently individual type III bursts actually result from multiple energy releases, and \cite{BCn:al-13} confirmed this idea by the Karl G. Jansky Very Large Array (VLA) observations with high temporal and spectral resolution.

\subsection{Association with jets and CMEs}
The association of the electron events in the present study with various types of EUV activities, including jets and large-scale mass motions, is in line with earlier studies on $^3$He-rich SEP events \citep{WanYM:al-06,Pic:al-06,Nit:al-06,Nitta2015}. However, we do not find a unique association with EUV jets. Y-M Wang and coworkers \citep{WanYM:al-06,Pic:al-06} used SoHO observations to identify EUV jets associated with $^3$He-rich SEP events (13/21 events with suitable observations). They suggested that EUV imaging with higher cadence would reveal more jets, and proposed a model where the EUV jet was the signature of magnetic reconnection between closed and open coronal magnetic field lines, which also led to the particle acceleration of electrons and ions. But high-cadence EUV imaging from STEREO and especially SDO used in the present work, as in \cite{Nitta2015}, does not confirm the expected systematic association between impulsive SEP events and coronal plasma jets. In some events the eruptive coronal activity may hide the plasma jet, but in some others jets are definitely not detected with SDO/AIA. However, there are also events with a clear timing correspondence, within a few minutes, of the electron acceleration in the corona revealed by hard X-ray and microwave emission, and the plasma jet. There is no a priori reason to exclude magnetic reconnection on the sole reason that no EUV jet is observed. The association with type III bursts and the PFSS extrapolations show the existence of open field lines in the parent active region. The scenario of interchange reconnection \citep[e.g.,][] {Shibata1994} as the origin of impulsive electron events therefore remains attractive, although the observations do not provide a unique simple picture where impulsive electron events would be exclusively associated with EUV jets, rather than with eruptive activity on larger scales.

The association of electron events with CMEs gives another hint to a more diversified picture of the origin of impulsive particle events than the historic two-class picture of either impulsive (flare-associated) or gradual (CME-associated) SEP-events \citep{Reames1999}. CMEs are not an occasional counterpart of the impulsive electron events studied here, but some white-light signature is observed with virtually all our electron events. In the only case where we did not identify a CME in the vicinity of the position angle of the parent flare, 2007 Jan 24, a faint signature might be hidden by a broad CME from a distinct active region. In the other events CMEs are seen in the corona. Their morphologies range from jet-like, which on occasion (2010 Nov 17) are clearly the extension of an EUV jet to the higher corona, to extended, as already reported by \cite{WanYM:al-06}. The combination of coronographic observations from SoHO and STEREO offered us the possibility to have in most events one spacecraft, which saw the parent activity close to the solar limb, so that the identification of a CME was much easier than in the earlier SoHO observations. The enhanced cadence of the STEREO coronagraphs also allows for a better timing identifications. In three out of 14 events where adequate coronographic observations were available, the CME came from previous erupting activity. They show that the CME is not a necessary condition for the electrons to achieve near-relativistic energies. But in nine cases the extrapolated height of the CME at the time of the hard X-ray and microwave burst, i.e. at the time of acceleration of near-relativistic electrons, was a few fractions of a solar radius above the photosphere. The presence of these low-altitude CMEs seems to be associated with those events at larger longitudinal separation angles (see Fig. \ref{fig:tdiff_longdiff}). These CMEs might hence play a role in widening the injection region, for example due to field-line spreading or deflection in the corona, or to electron acceleration on open field lines remote from the parent active region \citep[e.g.,][]{CSM:al-16}.

\section{Summary and conclusions}
For 17 different solar flare events we correlated the HXR spectral characteristics measured by RHESSI with the corresponding spectra of electron events observed in situ with STEREO/SEPT. Most of the in situ electron events show broken power-law spectra presumably caused by transport effects. At least two processes during interplanetary transport have been identified that are capable of causing such spectral breaks: i) the generation of Langmuir turbulence and ii) pitch-angle scattering (see discussion above). However, the range of the potential positions of these different breaks overlap at $\sim $\SI{100}{\keV}. This and the limited energy range and resolution of energetic particle instruments are likely the reasons why usually only a single break is identified in solar energetic electron spectra \cite[e.g., ][]{Lin1982, Reames1985, Krucker2009, Dresing2020}. It can therefore be not straightforward to identify which part of the spectrum is least influenced by the above effects and best suited to infer the acceleration spectrum.

We investigated the correlation of the HXR spectral index with both spectral indices (in case of broken-power-law shapes) observed in the in situ spectra. We find a good correlation of $\sim 0.8$ for both sets of value pairs with an alignment along the thin-target solution, i.e. a shift toward harder in situ electron spectral indices, when using the spectral index below the break $\delta_1$ or a shift toward the thick-target solution when using $\delta_2$.
All of our events would fall into the class of delayed events as defined by \cite{Krucker2007} who found only a low correlation of $\sim 0.4$ for these events. Although it cannot be ruled out that many of the observed onset delays are only apparent delays caused by instrumental effects, such as occasional poor pitch-angle coverage of STEREO/SEPT, the majority of our events are accompanied not only by EUV jets but also by CMEs. While the jets are likely a sign of interchange reconnection providing the flare-accelerated electrons with a connection to open field lines, a potential role of the CMEs in the acceleration process as suggested by \cite{Petrosian2016}, which could also cause the observed delays, cannot be ruled out. However, we do not find an effect of the CMEs on the correlation of the spectral indices. Nevertheless, the CMEs could have perturbed the transport of electrons through the lower corona, which might occasionally have caused a wider or shifted injection into interplanetary space as suggested by the correlation of CMEs with events observed at larger longitudinal separation angles.

We find clearly improving correlations when only considering events, which show significant anisotropies in the in situ electron observations. This suggests that transport effects such as pitch-angle scattering can reduce the spectral imprint of the acceleration and need to be taken into account when inferring the accelerated electron spectrum from spacecraft measurements.

Analysis of the starting frequencies of the associated type III radio bursts suggests that the acceleration height of most of our events was below 0.1 R$_\odot$ above the photosphere.
Furthermore, a detailed inspection of radio and microwave observations suggests that the electron fluxes at 1 AU could result from multiple injections of electrons in the corona as the majority of our events is accompanied by groups of type III  bursts. However, higher energy, i.e. near-relativistic electrons are only accelerated during part of these acceleration episodes as an indicated by the shorter hard X-ray and microwave emission.

in situ measurements of solar energetic electrons will remain a key observable to study acceleration, injection, and transport processes of SEP events. Furthermore, as their propagation time from the Sun to Earth is significantly shorter than that of associated solar energetic ions, one main application of solar energetic electrons is space weather forecast as used for example in the Relativistic Electron Alert System for Exploration \citep[REleASE,][]{Posner2009}.
Electron event observations during the upcoming solar cycle taken by new space missions such as Parker Solar Probe or Solar Orbiter open up new opportunities to understand solar energetic electrons events.
Much advantageous over the study presented here, Solar Orbiter will allow to detect the HXR flare and the in situ electrons at the same spacecraft, with the Spectrometer/Telescope for Imaging X-rays \citep[STIX;][]{Krucker2020} and the Energetic Particle Detector \citep[EPD;][]{Rodriguez-Pacheco2020}. Furthermore, both new space missions will take measurements at much smaller radial distance than 1 AU allowing to tackle the effect of interplanetary transport especially when combined with 1 AU baseline observations provided for instance by, STEREO~A, SOHO, ACE, and Wind. The new state-of-the art instruments also provide energetic electron measurements over a wider energy range with very fine energy resolution, which might finally allow to separate the imprints of acceleration and transport in the energy spectra.

\begin{acknowledgements}
 The work of A. W. was supported by DLR under grant No. 50 QL 1701. F.E. and N.D. acknowledge support from NASA grant NNX17AK25G and F.E. from DFG grant EF 98/4-1. N.D. acknowledges financial support by DLR under grant 50OC1702. We thank the International Space Science Institute (ISSI) for hosting our team on “Solar flare acceleration signatures and their connection to solar energetic particles.” L.G. acknowledges the NASA DRIVE SolFER Science Center grant 80NSSC20K0627. This study has received funding from the European Union’s Horizon 2020 research and innovation program under grant agreement No.\ 101004159 (SERPENTINE).
\end{acknowledgements}

\bibliographystyle{aa}
\bibliography{references}

\newpage
\begin{appendix}
\section{Additional information on the analyzed events}\label{appendix}
Table~\ref{tab:additional_info} provides additional information on the events under study. Column one lists the HXR date and peak time. Columns 2 and 3 list the characteristics of the associated microwave bursts: the start and end time, the highest frequency where the nonthermal microwave signature was observed, and the type of the predominant microwave emission, thermal (th), gyrosynchrotron (gs) or plasma emission (p). Column 4 lists the start and end time of the decametric type III burst, read from the dynamic spectra observed by Wind or STEREO near 10 MHz, using data with 1 min integration.
Column 5 indicates if the event was associated with a type II radio burst and provides its start and end time. Columns 6 and 7 provide the magnetic polarity of the associated active region in the corona and locally at the spacecraft (indicated by column 8) during the SEE event. The last column provides a rough classification for the strength of the observed electron anisotropy during the early phase of the electron events.
\\
The polarity of the magnetic field in the corona was inferred from the potential field source surface (PFSS) extrapolation of the SoHO/MDI measurements in the photosphere. The method of \cite{Scr:DeR-03} implemented in the PFSS tool of the Solarsoft package was used. The polarity in the Table is the one of the open field lines rooted in the flaring active region.
The in situ magnetic field polarity was determined by separating between inward our outward pointing magnetic field vectors as measured by STEREO/MAG \citep{Acuna2008} during the time of each electron event.
The table shows, that the in situ electron events were often observed in opposite polarity sectors than coronal polarity of the corresponding flare. We do, however, not see any influence of the polarity on other parameters such as the onset delays or the quality of the spectral correlations.\\
In two events (2009 Dec 22 and 2012 Apr 16) type II bursts were observed at meter wavelengths, without counterpart in the Wind/WAVES spectrum below 14 MHz. Only the first was reported by NOAA/SWPC.
On 2009 Dec 22 the type II burst started with the intense microwave emission and HXR burst, and it lasted a few minutes longer. DH type III bursts were observed by Wind/WAVES during its entire duration. On 2012 Apr 16 the type II emission occurred during the decay of the microwave and HXR burst. DH type III bursts accompanied the microwave/HXR burst, but not the type II burst. No clear difference of the energy spectra of the electrons was found when compared to the rest of the events. \\
For the anisotropy classification, presented in the last column of table~\ref{tab:additional_info}, we first determined the anisotropies using the method described by \cite{Bruedern2018}, and categorized anisotropies $A<1$ as small, $1>A>2$ as medium, and $A>2$ as large. Mixed classifications are provided for events with anisotropies close to the limiting values of 1 or 2. The asterisks in the anisotropy column mark events with limited or poor pitch-angle coverage, which can lead to an underestimation of the anisotropy.

\begin{table*}
\caption{Complementary radio, magnetic polarity and anisotropy information of the analyzed event}
\label{tab:additional_info}
\small
         \centering
\begin{tabular}{ccrccccccc}
        \hline
        \hline
            \noalign{\smallskip}
   HXR           & Microwave burst & $\nu_{\rm max}$       & DH III     & II             &  \multicolumn{2}{c}{Magnetic polarity} &  &  \\
     peak time & start - end           & [GHz]/   & start-end  &               & corona & in situ &  s/c & anisotropy \\
                     &                             &  type                          &               &              &             &           &        &                   \\
    \hline
2007/01/24 00:31:24 & 00:31:10-00:32:00 & 9/gs   & 00:29-00:33 & --  &  neg & mixed & A & small/medium\\
2007/01/24 00:31:24 &        --                     & --       &  --                 & --  &  neg & mixed & B & small \\
2007/01/24 05:16:09 & 05:15:50-05:16:30 & 5/gs   & 05:12-05:19 & -- & neg & mixed & B & small\tablefootmark{*} \\
2007/01/24 05:16:09 &    --                         &  --      & --                  & --  &  neg & mixed & A & small/medium \\
2009/12/22 04:56:10 & 04:53:00-04:57:00 & 17/gs & 04:52-05:04 & 04:57-05:03 &  pos & pos & B & large \\
2010/02/08 03:12:24 & 03:12:20–03:14:00 & 17/gs & 03:13-03:16 & --  &  neg & neg & B &  small \\
2010/11/12 03:53:41 & 03:53:25–03:56:00 &  1/p    & 03:45-03:54 & --  &  neg & neg & B &  small/medium\tablefootmark{*}\\
2010/11/12 08:01:54 & 08:01:00–08:04:50 & 15/gs & 07:55-08:11 & --  &  neg & neg & B &  medium/large \tablefootmark{*}\\
2010/11/17 04:36:44 & 04:36:30-04:37:10 &  3/gs   & 04:31-04:37 & --  &  pos & neg & B &  small/medium \tablefootmark{*} \\
2011/03/24 17:04:34 & 17:03:00-17:06:50 & 15/gs  & 16:53-17:16 & --  &  pos & neg & B &  large \\
2012/01/12 00:51:58 & $\sim$00:49–00:53 &  5/gs  & 00:49-00:53 & --  &  neg & pos & B & small\tablefootmark{*}\\
2012/03/25 00:27:54 & 00:27:05–00:28:30 &  9/gs   & 00:22-00:28 & --  &  neg & neg & B & small\tablefootmark{*}\\
2012/04/16 00:26:14 & $\sim$00:25-00:28 &  4/gs   & 00:24-00:30 & 00:29-00:40 &  uncertain & neg & B & small/medium\tablefootmark{*}\\
2012/05/07 03:21:58 & 03:21:00–03:26      & 17/gs  & 03:20-03:29 & --  &  neg & neg & B & medium\tablefootmark{*}\\
2012/06/27 12:36:00 & 12:36:00–12:36:30 &  9/gs   & 12:34-12:40 & --  &  neg & neg & B & medium/large \\
2012/06/28 02:12:24 & 02:12:00–02:13:30 &  2/p    & 02:10-02:14 & --   & neg & pos & B & medium\tablefootmark{*}\\
2012/07/01 07:14:44 & 07:14:40–07:14:50 &  9/gs  & 07:09-07:25 & --  &  neg & pos & B & large\\
2014/03/19 16:26:14 & 16:29:00–16:32    & 15/th   & 16:24-16:30 & --  &  uncertain & pos & B & large\\
2014/06/09 17:05:04 & 17:04:30–17:06:20 & 15/gs & 17:03-17:07 & --  &  uncertain & pos & B &medium\tablefootmark{*}\\
    \hline
\end{tabular}
\tablefoot{
   \tablefoottext{*}
   Limited or poor pitch-angle coverage in SEPT measurements due to non-nominal magnetic field configuration.
   }
\end{table*}

Many of the analyzed flares were accompanied by coronal jets seen in EUV. Table \ref{Tab_CMEsum} summarizes the observations of these jets using EUV images.
For the first 4 events, EUV observations from the STEREO-A EUVI \citep[][]{Wuelser2004} at 195 {\AA} were used,  because they provide the best available time cadence for these dates, which are before the launch of the  Solar Dynamic Observatory \citep[SDO,][]{Pesnell2012} mission in 2010. However, the time cadence (5 minutes in average) and spatial resolution are not sufficient to clearly confirm the absence of jets, for instance during the first event. Despite these instrumental limitations, jets and eruptions are still clearly found for the other three events.
For the later events, data from the Atmospheric Imaging Assembly \citep[AIA,][]{Lemen2012} on board SDO was available. AIA has a cadence of 12 second and provides images of the Sun in 7 EUV filters. The 304 {\AA} filter was used to look for coronal jets within one hour of the X-ray peak time of the flare, as jets are typically bright in this wavelength. In the 13 events observed in the SDO era, only two events are not associated with a coronal jet. Most of the observed jets are observed at the time and location of the flare, as reported in the third column of table \ref{Tab_CMEsum}. The events for which no jet was found, or with a jet delayed or at a different location in the active region than a flare, are not particularly associated with delayed SEP electron events.

Most of the analyzed flares were accompanied by coronal mass ejections (CMEs). Table \ref{Tab_CMEsum} summarizes the observations using the coronagraphs aboard SoHO \citep[LASCO/C2,][]{Brueckner1995} or STEREO \citep[SECCHI/COR1,][]{Howard2008}. In order to avoid spurious associations due to projection effects, we restricted the sample to events where the parent active region was within 30$^\circ$ of the limb. The third column displays the central meridian distance (CMD) for the chosen spacecraft. The fifth column gives the time of first appearance in the field of view of the coronograph, together with the estimated heliocentric distance as identified in the images\footnote{Movies provided by JHelioviewer or \url{https://cdaw.gsfc.nasa.gov/stereo/daily_movies/}}. Whenever possible a rough estimate of the speed in the plane of the sky is also provided, as well as the height of the CME front at the time of the HXR peak, which is inferred from linear back projection. Column six provides comments for specific events.

CMEs are found during all events where adequate coronographic observations are available. The relationship with the flare and electron events is not always clear, however. The 2007 Jan 24 events are not related with the partial halo CME observed at the time of the flare. Since the CME appeared in LASCO images at the east limb before the west limb, it likely originated from a region in the eastern solar hemisphere. No active region was on the eastern solar disk this day, but a candidate crossed the east limb three days later. We therefore conclude that the partial halo CME was a backside event. In three cases the CME was already high above the solar limb at the time of the flare (2012 Apr 16 and Jun 27, 2014 Mar 19). The starting frequencies of the type III bursts associated with these events show that the electron acceleration must have occurred at lower height. In nine events the extrapolated height of the CME front at the time of the HXR burst is within 1 R$_{\odot}$ above the photosphere, so that its early rise may have been related with the electron acceleration as traced by the HXR and radio emissions. The CMEs are in general not large, and some are reminiscent of jets.

\begin{table*}
  \centering
  \caption{Jets and CMEs possibly related with the electron events and flares.}
  \label{Tab_CMEsum}
\begin{tabular}{llllll}
\hline
\hline
 & \multicolumn{2}{c}{EUV jet}                & \multicolumn{3}{c}{Coronal mass ejection} \\
 HXR peak time     & Instr & First/                     & Instrument/  & Height-time evolution                                  & Comment \\
                              &         & Distance      &                           CMD [deg]  & (first/$r_0$/$V$/$r$(HXR))                         & \\
(1)                          & (2)   & (3)     & (4)  & (5)                      & (6)               \\
\hline
2007/01/24 00:31 & STA & -/- $^{[1]}$ $^{[2]}$  & SoHO W61 & 01:32/2.3 R$_\odot$/-/-                      & E backside \\
2007/01/24 05:16 & STA & 05:15/$<5''$           & SoHO W64 & 06:06/2.5 R$_\odot$/ -/-                     & E backside \\
2009/12/22 04:56 & STA & 04:55/$<5''$           & STB W114 & 05:11/1.7 R$_\odot$/470/1.0 R$_{\odot}$      &  \\
2010/02/08 03:12 & STA & 03:20/$<5''$ $^{[3]}$  & STB W71   & 03:41/1.6 R$_{\odot}$/200/1.1 R$_{\odot}$   & jet  \\
2010/11/12 03:53 & AIA & 04:05/$<5''$ $^{[4]}$  & STA  W86  & 03:55/1.7 R$_{\odot}$/580/1.6 R$_{\odot}$   & faint diffuse  \\
2010/11/12 08:02 & AIA & 08:55/$10''$           & STA  W83  & 08:05/? R$_{\odot}$/700/1.4 R$_{\odot}$     & faint diffuse   \\
2010/11/17 04:37 & AIA & flare/$<5''$           & STA  E105 & 04:45/1.8 R$_{\odot}$/930/1.2 R$_{\odot}$   & jet  \\
2011/03/24 17:05 & AIA & flare/$<5''$           & CMD $> 30^\circ$    &                                   & \\
2012/01/12 00:52 & AIA & 01:12/$<5''$ $^{[3]}$ $^{[4]}  $  & STB W92   & 01:01/1.7R$_{\odot}$/520/1.3 R$_{\odot}$  &   \\
2012/03/25 00:28 & AIA & -/- $^{[1]}$           & STB W93  & 00:26/1.9R$_{\odot}$/860/1.7 R$_{\odot}$            &       \\
2012/04/16 00:26 & AIA & -/- $^{[1]}$           & SoHO E89 & 00:36/3.5 R$_\odot$/$\sim 490$/3.1 R$_{\odot}$ & narrow \\
2012/05/07 03:22 & AIA & -/$<5''$ $^{[2]}$ $^{[4]}$        & CMD $> 30^\circ$ \\
2012/06/27 12:36 & AIA & flare/$20''$           & SoHO E64 & 12:10/3.2 R$_\odot$/$\sim 760$/4.3 R$_{\odot}$ & pre-existing \\
2012/06/28 03:22 & AIA & flare/$<5''$ $^{[2]}$  & CMD $> 30^\circ$  & &  \\
2012/07/01 07:15 & AIA & flare/$<5''$           & STB W105 & 07:25/2.2R$_{\odot}$ /700/1.6 R$_{\odot}$      &              \\
2014/03/19 16:26 & AIA & flare/$15''$           & SoHO E81 & 15:24/4 R$_\odot$/- /-                         & pre-existing  \\
2014/06/09 17:05 & AIA & 17:20/$50''$           & SoHO E90 & 17:24/2.8 R$_\odot$/$\sim 680$/1.7 R$_{\odot}$ &                 \\
\hline
\end{tabular}
\tablefoot{
Columns: (1) Date (see Table \ref{tab:events}),
(2)-(3): EUV jet (instrument (2), start time/distance from HXR emission site (3), );
(4)-(6): CME (instrument and central meridian distance of the flare (4), first detection/heliocentric distance/ speed in the plane of the sky (km s$^{-1}$)/extrapolated heliocentric distance at the HXR peak) (5), comment (6)

Comments in column (3): $[1]$ jet not detected in the data; $[2]$ numerous jets in this active regions over a few hours; $[3]$ possible filament eruption; $[4]$ faint ejection.
}
\end{table*}

\subsection{Implications of the spectral variation in the two multi-spacecraft events of 24 Jan 2007}\label{app:multi-sc-events}
Two events on 24 Jan 2007 were observed by both STEREO spacecraft when they were still separated by less than one degree in longitude and latitude. As expected the peak intensities and onset times are very similar as detected by the two observers. However, both events were observed during changing magnetic field conditions leading to variable and partly nonoptimal pitch-angle coverage of the SEPT instruments. The pitch-angle coverage was always better at STEREO~A, which might be the reason for the slightly larger anisotropies compared to STEREO~B. For the first event, the determined spectral indices of the peak spectra are similar at the two spacecraft but for the second event STEREO~B only observes a single power law while STEREO~A detects a broken power law. Although the spectral values observed at the two spacecraft agree within their uncertainties for the two events this illustrates how the magnetic field configuration and variation can not only influence onset determinations but also the determined spectra.
It is expected that the most reliable spectra are those where the instruments detect the electron population propagating anti-sunward along the magnetic field, i.e. at pitch angle 0 or 180 (depending on the magnetic field polarity). If these directions are not covered by the instrument or if strong scattering has led to a vanished anisotropy, the determined spectra may carry systematic changes.
Unfortunately, periods of non-ideal pitch-angle coverage occur regularly in SEPT measurements so that several of the analyzed events may be subject to this limitation.

\subsection{The events on 17 Nov 2010, 28 June 2012, and 19 Mar 2014}\label{app:three-special-events}
An important issue discussed in this manuscript is that it is not straightforward to choose the appropriate spectral index out of the usually observed double power law spectrum for a comparison with its solar counterpart. Observations of near-relativistic electrons usually show only one spectral break or transition (e.g., using Wind/3DP \citep{Krucker2007} or STEREO/SEPT \citep{Dresing2020}). However, due to the limited overall energy range and energy resolution of these instruments their ability to resolve which of the different effects caused the spectral break or if an overlap of effects determines the spectral shape is also limited.
The correlation with the solar counterpart spectral index may even help here: When correlating the photon spectral index $\gamma$ with both the lower $\delta_1$ and upper $\delta_1$ spectral indices of the in situ electron spectra (Fig. \ref{fig:delta1_gamma} \&  \ref{fig:delta2_gamma}) we find that three events (17 Nov 2010, 28 June 2012, and 19 Mar 2014) do not fit the rest of the distributions. As described in section \ref{sec:analysis} we suspect that these three events show a spectral transition, which can rather be attributed to Langmuir-wave generation while the spectral transitions of the rest of the events are likely caused by pitch-angle scattering \citep[][]{Dresing2020, Strauss2020}. Indeed, when treating these three events accordingly in the correlation plots (Fig. \ref{fig:delta1_gamma} \&  \ref{fig:delta2_gamma}) the overall correlations increase significantly. One of these three events (17 Nov 2010) is indeed the event with the lowest break energy in our whole sample ($E_b=69$ keV), which supports the assumption that this break can be attributed to Langmuir-wave generation, which is expected to yield a break around 60 keV \citep{Krucker2009}.
The break energies of the other two special events (90 and 101 keV) are rather low with respect to the mean spectral break value of 120 keV found by \cite{Dresing2020}, which was attributed to pitch-angle scattering, however many other events analyzed here show similar low break energies.
The two latter events are both anisotropic. While the pitch-angle coverage during the 28 June 2012 event is not optimal, likely leading to an underestimation of the anisotropy, the pitch-angle coverage during the 19 Mar 2014 event is ideal and shows a very strong anisotropy (not shown). Consultation of the STEREO level3 Interplanetary Coronal Mass Ejection (ICME) list\footnote{\url{https://stereo-ssc.nascom.nasa.gov/data/ins_data/impact/level3/STEREO_Level3_ICME.pdf}} reveals that STEREO~B was embedded inside an ICME when the 19 Mar event occurred. The usually very quiet magnetic field conditions inside ICMEs may have contributed to very weak scattering conditions during this event and consequently high anisotropy.
The pitch-angle distribution during the 28 June event is peculiar as the highest intensities are observed in SEPT's north and ani-sun sectors, suggesting that also during this event, the spacecraft was embedded inside a non-nominal magnetic field configuration. However, an ICME passage is only reported for about six hours after the event onset. The solar origin of this electron event is, however, not doubted given the good temporal correlation with the flare and the associated type III radio burst, and the notable anisotropy.
In spite of the peculiarities of these two latter events, they are otherwise not outstanding when comparing their characteristics such as peak intensities or energies, strength of the anisotropy, onset delays or radio features with the rest of our sample. The reason why the spectral break due to Langmuir-wave generation is dominant in the spectra of these events, is therefore not conclusively understood.

\end{appendix}

\end{document}